\documentclass[aps,12pt,tightenlines,amsmath,amssymb]{revtex4}

\usepackage{graphicx}
\usepackage{dcolumn}
\usepackage{bm}
\usepackage{natbib}

\newcommand{\eps}{\varepsilon}
\newcommand{\sign}{\mathop{\rm sign}\nolimits}
\def\Np{{N+1}}
\def\one{{\bm1}}
\def\zero{{\bm0}}
\def\realnumbers{\mathbb R}

\def\Eqref#1{Eq.~(\ref{#1})}
\def\parenref#1{(\ref{#1})}

\def\cc#1#2{C^{#1}_{#2}}
\def\u#1#2{u_{#2}^{#1}}
\def\v#1#2{v_{#2}^{#1}}
\def\uu#1#2#3{u_{#2,#3}^{#1}}
\def\vv#1#2#3{v_{#2,#3}^{#1}}

\def\VNplusinv{V_{N+1}^{-1}}

\newcommand{\labeld}[1]{ }

\def\lsmatrix#1#2{\left(\begin{array}{#1}#2\end{array}\right)}
\def\lsmatrixcurved#1#2{\left(\begin{array}{#1}#2\end{array}\right)}

\def\lsarray#1#2{\begin{array}{#1}#2\end{array}}
\def\lsdet#1#2{\left|\begin{array}{#1}#2\end{array}\right|}
\newcommand{\be}{\begin{equation}}\newcommand{\ee}{\end{equation}}
\newcommand{\bea}{\begin{eqnarray}}\newcommand{\eea}{\end{eqnarray}}
\newcommand{\bean}{\begin{eqnarray*}}\newcommand{\eean}{\end{eqnarray*}}

\def\ll{lower-left}

\renewcommand{\d}{{\rm d}}
\newcommand{\der}[1]{\frac{\d#1}{\d t}}
\newcommand{\e}{{\rm e}}
\newcommand{\half}{{\textstyle{\frac12}}}
\newcommand{\mean}[1]{\left\langle#1\right\rangle}
\newcommand{\quarter}{{\textstyle{\frac14}}}
\newcommand{\s}{\sigma}
\newcommand{\tr}{\mathop{\rm tr}\nolimits}
\newcommand{\A}{{\mathcal{A}}}
\newcommand{\B}{{\mathcal{B}}}
\newcommand{\C}{{\mathrm{C}}}
\renewcommand{\Im}{\mathop{\rm Im}\nolimits}

\newcounter{remark}

\begin{document}
\title{Spectral properties of zero temperature dynamics in a model of a compacting granular column}
\author{L. S. Schulman}
\affiliation{Physics Department, Clarkson University, Potsdam, New York
13699-5820, USA}
\email{schulman@clarkson.edu}
\author{J. M. Luck}
\affiliation{Institut de Physique Th\'eorique, IPhT, CEA Saclay and URA 2306, CNRS, 91191 Gif-sur-Yvette cedex, France}
\email{jean-marc.luck@cea.fr}
\author{Anita Mehta}
\affiliation{Theory Department, S. N. Bose National Centre, Block JD Sector III, Salt Lake, Calcutta 700098, India}
\email{anita@bose.res.in}

\begin{abstract}
The compacting of a column of grains has been studied using a one-dimensional Ising model with long range directed interactions in which down and up spins represent orientations of the grain having or not having an associated void. When the column is not shaken (zero ``temperature'') the motion becomes highly constrained and under most circumstances we find that the generator of the stochastic dynamics assumes an unusual form: many eigenvalues become degenerate, but the associated multi-dimensional invariant spaces have but a single eigenvector. There is no spectral expansion and a Jordan form must be used. Many properties of the dynamics are established here analytically; some are not. General issues associated with the Jordan form are also taken up.
\end{abstract}

\date{\today}
\maketitle

\medskip

\section{Introduction\label{sec:intro} \labeld{sec:intro}}

Zero temperature dynamics is parsimonious. It is dominated by its constraints and lends itself to non-generic behavior. Our focus is on a model of granular compaction and we will see that the peculiarities of the allowed motions induce non-exponential relaxation even when all eigenvalues for the stochastic dynamics are real. The mathematical mechanism behind this uncharacteristic behavior for a Markov process lies in the fact that the associated eigenvectors do not span the entire state space and the best one can do is to represent the generator of the stochastic dynamics as a Jordan form. Although it is rare for this mathematical construct to appear in a physical context, it is not unheard of \cite{poincaresemigroup, daems, ocinneide, grassmann, dhar, stephen, sadhudhar}, and in particular, in some of these there are also instances of non-exponential decay despite the absence of memory.

Granular materials can exhibit features significantly different from those of traditional fluids \cite{mehtabarkerluckphysicstoday}. A model of such materials was developed in \cite{mehtaluck4, luckmehta1, luckmehta2, luckmehta3, luckmehtapnas} and also includes (strictly) positive temperature, which in this context relates to random shaking of the material. The model~\cite{mehtaluck4, luckmehta1} consists of a finite column of $N$ grains, labeled by their depth $n=1,\dots,N$\@. Each grain has an orientation variable $\sigma_n=\pm1$\@. Grain $n$ is called \textit{up} or \textit{ordered} when $\sigma_n=+1$ and \textit{down} or \textit{disordered} when $\sigma_n=-1$\@. An ordered grain occupies one space unit. A disordered one traps a void and wastes space; it is said to occupy space $\eps$, so that it traps a void  of size $(1-\eps$) alongside it. The quantity $\eps$ is effectively a shape parameter; see \cite{luckmehtapnas, mehtabarkerluckphysicstoday}. Within the column we define a \textit{local field}, $h_n$,  whose purpose is to model the effect of compacting constraints. In the simplest case \cite{mehtaluck4, luckmehta1}, this is only due to grains \textit{above} grain-$n$:
\be
h_n=\sum_{m=1}^{n-1}f(\sigma_m) =\eps n_0 -n_1 \,,
   \hbox{~with the ``shape factor'' }f(\sigma)=\begin{cases}-1,&\sigma=+1\\ ~\eps,&\sigma=-1\end{cases}
\label{eq:hfield}
\,.
\ee
\labeld{eq:hfield}
In \Eqref{eq:hfield}, $n_1$ is the number of plus spins above spin-$n$, and $n_0$ the number of minus spins above it. Under the dynamics (to be specified in a moment) spins \textit{tend to} order in the field direction at finite temperature and \textit{must} orient along the field at zero temperature. This models the observed local compaction in granular materials: the system tends to eliminate its voids~\cite{barkermehta, bergmehta}.

The system undergoes continuous Markov dynamics generated by a matrix of transition probabilities. In one unit of physical time the system averages $N$ steps. Zero-temperature dynamics tends to retrieve ground states, and is defined as~\cite{mehtaluck4, luckmehta1}:
\be
\sigma_n\to\sign{h}_n \,.
\ee
For vanishing $h_n$, the simplest version of the dynamics gives a 50\% chance of a switch in $\sigma_n$\@. Surprisingly perhaps, all $2^N$ configurations can be reached, implying irreducibility of the transition matrix and strict positivity of the stationary state.

Although our preliminary work using the observable representation \cite{multiplephases, meanfieldobsrep, imaging} has proved useful for non-zero temperature dynamics in this model, it turns out that there is a serious obstacle to its application when the constraints of zero temperature are imposed: the stochastic matrix generated under these rules \textit{does not have a full complement of eigenvectors}.

In this article we will develop in detail the stochastic matrix governing this process and demonstrate the aforesaid properties. Much of the power of our proofs arises from the use of a convenient basis. The eigenvalue spectrum turns out to be extremely simple and is shared by other models. We will show that appropriate combinatorial coefficients characterize the dimension of the space associated with each eigenvalue. Moreover, in some cases there is even a finer structure when, modifying the rule just given, the $h=0$ rate is changed from $\half$ to $\half+\delta$ for (strictly) positive or negative $\delta$ (but with $|\delta|<\half$). Here too the dimension associated with each eigenvalue is given by a combinatorial coefficient.

Nevertheless, there is still a great deal that is unknown in this system. For example, we can characterize basis vectors for only half the dimensions of the various invariant spaces (the construct that replaces eigenvectors). This is related to what we consider to be one of the important questions raised here. In quantum mechanics symmetries usually relate degenerate states. What is it that unites states in the same invariant space?  Another question, not unique to this work, is, why the Jordan form? It is non-generic; in fact we will display results showing that even at zero temperature a slight change---in particular the non-zero $\delta$ modification mentioned above, but at different $\eps$ values---returns us to the conventional world of successful spectral decomposition. For both these questions we will present intuitions backed by mathematics, but ---alas---not complete proofs. For example, the invariant spaces turn out to have dimensions given by combinatorial coefficients, and these will be shown to arise because of the ways of choosing a given number of spins out of the entire collection. So the dimension is accounted for, but there are details of the characterization of the space that are missing. Similarly, the Jordan form will reflect a cascade process in decay, as the system finds its way to the stationary state in the face of the many constraints imposed by zero-temperature dynamics. We are able to prove that a Jordan form is needed, i.e., that the eigenvectors cannot span. But as to showing that each invariant space has but a single eigenvector, we come tantalizingly close, but gaps remain.

In Sec.\ \ref{sec:stochasticdynamics} we present the matrix generator of the stochastic dynamics that implements the rules just given. With appropriate numbering of its states it satisfies a recursion relation as the number of spins is increased. That recursion is sufficient to deduce the spectrum of the dynamics and to allow generalization to other similarly structured matrices. The multiplicity of invariant spaces is also established. In this section we also introduce a reference model, a random walk on the edges of a hypercube, that plays a role later in our development. In Sec.\ \ref{sec:eigenvectors} we examine the eigenvectors---and non-eigenvectors---associated with some of the eigenvalues. Following that, in Sec.\ \ref{sec:timedependence} the anomalous time-dependence is explored. In Sec.\ \ref{sec:spincorrelations} we approach the dynamics from a different perspective and study the behavior of correlations, which because of the quasi-Boolean nature of our variables is equivalent to the analysis of probabilities. Using the independence of spin-$k$ dynamics on that of spins below it, we establish that for the correlations the dynamical equations assume a triangular form. This provides another way of discerning the spectrum and other properties, and gives insight into the physical basis of the properties derived from the abstract algebraic approach. In Sec.\ \ref{sec:dyn2s} we work through the smallest non-trivial example to illustrate our general results.

It also turns out that the zero-temperature dynamics does not always lead to a Jordan form. In particular, when the field $h$ can vanish in the interior (possible for particular values of the parameter $\epsilon$) and for $\delta\neq0$, the generator of the stochastic dynamics acquires complex spectrum, with a full set of eigenfunctions. This material is in Sec.~\ref{sec:nongeneric}. Following that, in Sec.\ \ref{sec:sumrule} we take another perspective and deduce further dynamical properties based on the existence of a sum rule for decay rates.

Finally, in Sec.\ \ref{sec:triangular} we show that the basis for diagonalizing the hypercube random walk brings the stochastic dynamics matrix to triangular form thereby unifying the two approaches, that of correlations and that of algebraic recursions. As for the other recursion results, this triangularity is more general than the particular physical model from which we began.

The last section reviews and discusses our results.

\section{The stochastic dynamics \label{sec:stochasticdynamics} \labeld{sec:stochasticdynamics}}

\subsection{Defining and labeling the generating matrix\label{sec:w} \labeld{sec:w}}

The central object of study is the generator of the stochastic dynamics. For convenience we change state-label conventions from that in Refs.\ \cite{luckmehta1, luckmehtapnas, luckmehta2}. Let the state of spin $k$ counting from the top be $\mu_k\equiv(\sigma_k+1)/2$\@. The binary string of 0's and 1's corresponding to an $N$-spin state must be ordered when the transition probabilities are written in matrix form. For reasons that will become evident we label states in \textit{reverse} binary order. That is $\ell_\mu=1+\sum_{k=1}^N \mu_k 2^{k-1}$, with $\mu=(\mu_1,\dots,\mu_N)$ and $\mu_1$ the \textit{top} spin. Here are two examples, $N=2$ and $N=3$:
\be
\begin {array}{cccc}
\ell_\mu &~&\mu_1&\mu_2\\
1&~&0&0\\
2&&1&0\\
3&&0&1\\
4&&1&1
\end {array}
\quad\,,\qquad\qquad
\begin {array}{ccccc}
\ell_\mu &~&\mu_1&\mu_2&\mu_3\\
1&~&0&0&0\\
2&~&1&0&0\\
3&~&0&1&0\\
4&~&1&1&0\\
5&~&0&0&1\\
6&~&1&0&1\\
7&~&0&1&1\\
8&~&1&1&1
\end {array}
\ee

Following the rules outlined above, we give the transition probability for going from state-$\mu$ to state-$\nu$\@. This rate will be called $w(\nu,\mu)$ and is the probability, per unit microscopic time, for the transition $\nu\leftarrow\mu$ (so one reads from right to left). It is the \textit{continuous} time generator and its diagonal is adjusted so that $\sum_\nu w(\nu,\mu)=0$\@. The fact that column sums of $w$ add to zero already implies that it has the left eigenvector $A_0(\nu)\equiv1$, with eigenvalue 0. The corresponding right eigenvector is the stationary state, which by virtue of the irreducibility mentioned earlier is non-degenerate and strictly positive \cite{note:donotneedproof}. By a variation of Perron-Frobenius theory, it is known that all eigenvalues of $w$ have 0 or negative real parts. Entries in $w$ will be referred to either by giving $\mu$ or $\ell_\mu$ as defined above. Time evolution acts to the right on probability distributions (``$p$'') in the following way: $p(t)=\exp(wt)p(0)$\@.

For $\mu\neq\nu$, the $2^{N}\!\!\times\!2^{N}$ matrix $w(\nu,\mu)$ (or $w_N(\nu,\mu)$ when we wish to emphasize its $N$-dependence) is zero unless the states $\mu$ and $\nu$ differ in precisely one spin entry. Let that spin be the $m^\mathrm{th}$ (from the top). Let the ``change'' in going from $\mu$ to $\nu$ be defined as the binary value of the target ($\nu_m$) in the $m^\mathrm{th}$ site minus the binary value of the source ($\mu_m$) at that site. Recall that the field at level $n$ is defined as $h_n=\eps n_0 -n_1$, with $n_0$ the number of down spins (strictly) above spin $n$ and $n_1$ the number of up spins. $\eps$ is a parameter. Note that the labels 0 and 1 on the $n$'s in the definition of $h$ now correspond to the $\mu_n$ value. If the field and the ``change'' are both positive, or both negative, the transition can take place and $w_N(\nu,\mu)=1$, otherwise not (in which case $w_N(\nu,\mu)=0$). If the field is zero, $w_N(\nu,\mu)=1/2+\delta$, irrespective of the change, with $\delta$ a second parameter.

For any particular $N$, the set of $\eps$ values breaks into two classes: ``generic'' and ``non-generic'' or ``special'' \cite{note:rational, luckmehta1}. ``Generic'' means the only place the field $h$ can be zero is above the first spin. This corresponds to irrational $\eps$ or to rational numbers that (in reduced form) involve sufficiently large integers. ``Non-generic'' $\eps$ allows the field to vanish within the column and leads to enhanced fluctuations and to significant $\delta$-dependent features in the spectrum for non-zero $\delta$\@. Because in our zero-temperature model being exactly zero is different from being almost zero, $w$ is stepwise constant as a function of $\eps$ and its value at ``special'' $\eps$'s is not its limit as $\eps$ approaches these values.

\subsection{Recursion and eigenvalue spectrum\label{sec:recursion}\labeld{sec:recursion}}

The most important features of $w_N$ arise from its recursive structure as a function of $N$, most evident in the state enumeration listed above. We can start with $N=1$, $\delta=0$, for which our assertion that
\def\half{\frac{1}{2}}
\be
w_1=\lsmatrix{rr}{-\half&\half\\ \noalign{\vskip1mm} \half&-\half}
\label{eq:w1}
\,,
\ee
\labeld{eq:w1}
requires a one sentence justification: the field, $h$, is zero, putting $\half$'s on both off-diagonals, and the diagonal is adjusted to give zero column sums. For $N=2$ the uppermost spin again sees zero field so that the upper-left and lower-right blocks are the same (i.e., as in \Eqref{eq:w1}), except for the diagonal. This is a consequence of the numbering scheme. In particular the $\mu_1$ values have the same pattern in both blocks; they differ only in their $\mu_2$ value.

The off-diagonal blocks (both 2-by-2) deal with transitions in which \textit{only} the deepest spin (in this case number 2) is changed. Hence it can only have entries on its diagonal. These entries are 0, 1 or $1/2$ depending on the sign of the field or whether it's zero. Before correcting the diagonal for zero column sum, $w_2$ has the following appearance:
\be
(w_2)_{\hbox{\tiny non-diagonal portion}}=\lsmatrix{cc}
              {(w_1)_{\hbox{\tiny non-diagonal portion}} & \Delta\\
              \noalign{\vskip3pt}
               \widetilde\Delta&(w_1)_{\hbox{\tiny non-diagonal portion}}}
\,,
\label{eq:recursionsansdiagonal}
\ee
\labeld{eq:recursionsansdiagonal}
where $\Delta$ and $\widetilde\Delta$ are themselves diagonal matrices (of the same size as $w_1$). What is important to note is that $\Delta+\widetilde\Delta=\one$, the size-$w_1$ identity matrix. This is because for each pair of states on $\Delta$'s diagonal (which differ in a single spin) if the field is positive for $\Delta$ it is negative for $\widetilde\Delta$ so one matrix element is 1, the other zero. If the field is zero, it is zero for both, and both are $1/2$\@. A moment's reflection shows that getting the column sums right is also easy and the form of $w_2$ is
\be
w_2=\lsmatrix{cc}
              {w_1-\widetilde\Delta    &      \Delta\\
               \widetilde\Delta        &  w_1-\Delta}
\,,
\label{eq:recursionforw2}
\ee
\labeld{eq:recursionforw2}
The exact form of $\Delta$---which is to say, where it has zeros, ones and halves---will depend on $\eps$\@. (In Sec.\ \ref{sec:nongeneric} we take up the non-zero $\delta$ case, which for non-generic $\eps$ can violate $\Delta+\widetilde\Delta=\one$\@.)

The arguments we have just given for building $w_2$ from $w_1$ are valid for any $N$ and by the same reasoning one has
\be
w_{N+1}=\lsmatrix{cc}
              {w_N-\widetilde\Delta_N   &     \Delta_N\\
               \widetilde\Delta_N       &  w_N-\Delta_N}
\,.
\label{eq:recursion}
\ee
\labeld{eq:recursion}
Here we have added the label $N$ to $\Delta$ indicating that it is a $2^N$-by-$2^N$ matrix, as is $w_N$\@. (A word of caution on notation: $\Delta_N$ enters the $(N+1)$-spin dynamics; it is \textit{not} part of $w_N$.) Moreover,
\be
\Delta_N+\widetilde\Delta_N={\one}_N
\,,
\label{eq:sumofDeltas}
\ee
\labeld{eq:sumofDeltas}
for $\delta=0$ or for non-zero $\delta$ in the case of generic $\eps$, since the field cannot vanish within the column. (Note that ${\one}_N$ is not the $N$-by-$N$ identity, but the $2^N$-by-$2^N$ identity. Similarly $\zero_N$ is the $2^N$-by-$2^N$ matrix of zeros.)

It is remarkable that the full eigenvalue spectrum as well as the dimensions of the invariant spaces (whether a Jordan form is needed or not) can be deduced from this recursion alone (plus properties of $w_1$ that allow an induction). The argument proceeds by examining the characteristic polynomial of $w_N$\@. Let
\be
P_N(\lambda)\equiv\det\left(w_N-\lambda \one_N\right)
\,.
\ee
Then:
\bea
P_\Np(\lambda)&\equiv&
\lsdet{cc}{w_N-\tilde\Delta -\lambda \one_N& \Delta \\ \tilde \Delta & w_N-\Delta-\lambda \one_N}
\label{eq:det-step0}\\
&=&\lsdet{cc}{w_N -\lambda \one_N& w_N-\lambda \one_N \\ \tilde \Delta & w_N-\Delta-\lambda \one_N}
\label{eq:det-step1}\\
&=&\lsdet{cc}{w_N -\lambda \one_N& \zero_N \\ \tilde\Delta & w_N-\Delta-\tilde\Delta-\lambda \one_N}
\label{eq:det-step2}\\
&=&\lsdet{cc}{w_N -\lambda \one_N& \zero_N \\ \tilde\Delta & w_N-(\lambda+1) \one_N}
\label{eq:det-step3}\\
\noalign{\smallskip}
&=& \left|\,w_N -\lambda \one_N\right|\cdot \left|\,w_N -(\lambda+1) \one_N\right|
\,.
\label{eq:detmanipulations}
\eea
The step from \Eqref{eq:det-step2} to \Eqref{eq:det-step3} depends on \Eqref{eq:sumofDeltas}. (As indicated, this holds for generic $\epsilon$ and any $\delta$, or for non-generic $\epsilon$ and zero $\delta$\@. It does \textit{not} hold for non-generic $\epsilon$ and non-zero $\delta$\@. For that case see Sec.~\ref{sec:nongeneric}.)

The inductive hypothesis is
\be
P_N(\lambda)=\prod_{k=0}^N \left(\lambda+k\right)^{\cc N k}
\,,
\ee
where $\cc Nk=N!/k!(N-k)!$, the combinatorial coefficient. This assertion is trivial for $N=1$ or 2.

From \Eqref{eq:detmanipulations} it follows that
\bea
P_\Np&=&\prod_{k=0}^N \left(\lambda+k\right)^{\cc N k}\prod_{k=0}^N \left(\lambda+k+1\right)^{\cc N k}\\
&=&\lambda(\lambda+N+1)\prod_{k=1}^N \left(\lambda+k\right)^{\cc N k + \cc N {k-1}}
\,.\eea
The inductive hypothesis is now proved by observing that
\be
\cc{N+1}k=\cc N k + \cc N {k-1}
\label{eq:combinatorialidentity}
\,.\ee
It follows that the eigenvalues are $\{0,-1,\dots,-N\}$, with eigenvalue $-k$ having an invariant space of multiplicity~$\cc{N}k$\@.

Because of our analytic information on the spectrum and multiplicities, this situation will be referred to as ``integrable.'' This is to be contrasted with the non-generic $\eps$, non-zero $\delta$ case, which we will characterize as ``non-integrable.''

\refstepcounter{remark} \label{rem:y}
\smallskip\noindent\textsf{Remark \arabic{remark}}:~
\labeld{rem:y} The spectrum and invariant space structure are the same for any family of matrices obeying Eqs.\ (\ref{eq:recursion}) and (\ref{eq:sumofDeltas}) for which the induction can be initiated. An illustrative example is $\Delta_k=\half{\one}_k$ for $k\leq N-1$\@. Call the associated matrix $y_N$\@. This is the generator of a random walk on the $N$-cube. It can also be looked upon as the infinite temperature limit of an $N$-spin Ising model in which only single spin flips are allowed. Unlike our zero-temperature granular dynamics, the matrix $y$ is symmetric and has a full complement of eigenvectors. They are given by the following construction. Let $A$ be a subset of the numbers 1 through $N$\@. There are $2^N$ such subsets. For this subset define a particular $2^N$-vector, $v_A$, as follows. Let $\mu=(\mu_1,\dots,\mu_N)$ be an $N$-string of 0's and 1's. Then $v_A(\mu)=(-1)^{n(\mu,A)}$, where $n(\mu,A)$ is the number of 1's in $\mu$ that fall in the subset $A$\@. (In other words, if $\chi_{_A}$ is the characteristic function of $A$, $n(\mu,A)=\sum_k\mu_k\cdot\chi_{_A}(k)$\@.) It is not difficult to show that this is indeed an eigenvector of $y_N$ and has eigenvalue $-|A|$, with vertical bars indicating cardinality. This immediately implies that the multiplicity of the eigenvalue $-k$ is~$\cc N k$\@. \`A propos the $N$-cube interpretation, when the image of this random walk under the observable representation \cite{note:OR} is plotted, one gets in fact a cube, not at all surprising in view of~\cite{imaging}.

\refstepcounter{remark} \label{rem:isospectralone}
\smallskip\noindent\textsf{Remark \arabic{remark}}:~
\labeld{rem:isospectralone} Many other matrices having the same eigenvalues and invariant space dimensions can be constructed by varying $\Delta_k$ as you build toward $N$\@. Some require the Jordan form, some do not. There is no need to have the entries in $\Delta$ be confined to $\{0,1/2,1\}$, nor even to the reals. There is no need to start from the $w_1$ given above. For example one can start from
\begin{enumerate}
\item\label{item:startinduction1} $w_0=0$, a 1-by-1 matrix. It has the single eigenvalue 0.
\item\label{item:startinduction2} $w_1=\left(\begin{array}{ccc}a&~&\mu(a+1) \\ -a/\mu &&-(a+1)  \end{array}\right) $\@. This has eigenvalues 0 and 1\@. It is only stochastic for $\mu=1$\@.
\end{enumerate}
So there is also no requirement that the matrix be stochastic. The matrix $\Delta$ can be anything of appropriate size. Possibility \ref{item:startinduction2} is more general than \ref{item:startinduction1}\@. The induction can also be started at larger matrices, say, $w_2$, allowing yet larger classes of isospectral operators. Approach \ref{item:startinduction2} reduces to Approach \ref{item:startinduction1} for $\mu=1$, $a=-1/2$\@.

\refstepcounter{remark} \label{rem:maximaljordan}
\smallskip\noindent\textsf{Remark \arabic{remark}}:~
\labeld{rem:maximaljordan} In our numerical experience, \textit{all} $N>1$ matrices constructed under the rules for granular dynamics were maximally Jordan, by which we mean that there are degenerate eigenvalues and that each invariant space has but a single eigenvector. (This includes generic $\eps$, $\delta\neq0$, but does not apply to the non-integrable situation, i.e., non-generic $\eps$, non-zero $\delta$, where the eigenvalues cease to be degenerate.)

\refstepcounter{remark} \label{rem:conjugation}
\smallskip\noindent\textsf{Remark \arabic{remark}}:~
\labeld{rem:conjugation}
The same operations by which we step-by-step simplified the determinant can be performed on the original matrix, casting additional light on its structure. See Appendix~\ref{sec:conjugation}.

\refstepcounter{remark} \label{rem:isospectraltwo}
\smallskip\noindent\textsf{Remark \arabic{remark}}:~
\labeld{rem:isospectraltwo}
In building matrices using the recursion \Eqref{eq:recursion}, if the entries of the diagonal matrix $\Delta$ are taken randomly from the set $\{0,\frac12,1\}$, one often has non-spanning eigenvectors, i.e., the need for Jordan forms. However, the invariant spaces are not necessarily maximal, i.e., there can be more than one eigenvector with given eigenvalue. We further remark that discerning whether or not a matrix requires a Jordan form can be numerically delicate. One indicator is that when one attempts to diagonalize by conventional methods one finds that the associated ``eigenvalues'' contradict analytically determined properties, for example, in having a non-zero imaginary part (see Sec.\ \ref{sec:falsediag} for the explanation).

\bigskip

We next consider non-zero $\delta$, generic $\epsilon$, so that the only site that can experience zero-field is the top one. The only transitions affected are those between an odd-numbered site and the even-numbered site with index one larger.  Then the change in $w$ due to the presence of $\delta$ is
\be
\frac{\d w_{\!N}}{\d\delta}=
\left(
\begin{array}{ccccc}
\sigma & 0      & 0      &  &\dots\\
0      & \sigma & 0      &  &\dots\\
\vdots &        &\ddots  &  & \\
0      & \dots  &        & &\sigma
\end{array}
\right)
\hbox{~with~} \sigma \equiv \left(\begin{array}{rr} -1&1\\1&-1 \end{array} \right)
\,.
\ee
There are $2^{N-1}$ copies of $\sigma$ in $w_N$\@. Since $w$ is linear in $\delta$, $w_N(\delta)=w_N(0)+\delta\frac{\d w_{\!N}}{\d\delta}$\@. For $P_N(\lambda,\delta)$ (defined as $\det\left(w_N(\delta)-\lambda I_N\right)$) we make the inductive hypothesis:
\be
P_N(\lambda,\delta)=\prod_{k=0}^N \left(\lambda+k\right)^{\cc {N-1} k}
          \left(\lambda+k+2\delta\right)^{\cc {N-1} {k-1}}
\label{eq:spectrumwithdelta}
\,.
\ee
\labeld{eq:algebraicspectrum}
(Recall (\Eqref{eq:combinatorialidentity}) that ${\cc {N} {k}}={\cc {N-1} k}+{\cc {N-1} {k-1}}$, so for $\delta=0$ this reduces to the former case.) This hypothesis can be directly verified for $N=2$ or 3, and follows from our Eqs.\ (\ref{eq:det-step0})--(\ref{eq:detmanipulations}) (which, as indicated, remain true for generic $\epsilon$) together with identities of the form \Eqref{eq:combinatorialidentity}.  In Sec.\ \ref{sec:extensions} we provide another way of reaching the same conclusions.

It follows that the invariant subspaces are now smaller: new eigenvectors emerge, one per invariant space. The span of the $\delta=0$ invariant space associated with the eigenvalue $-k$ is \textit{not} the sum of those for eigenvalues $-k$ and $-k-2\delta$ for non-zero~$\delta$\@.

\section{Eigenvectors and invariant spaces\label{sec:eigenvectors}\labeld{sec:eigenvectors}}

To address the nature of the eigenvectors and invariant spaces additional tools will be used. It will also be useful to work with the transpose of $w$, since its \textit{right} eigenvectors often have simpler structure. Let $g_N\equiv w_N^\top$ (with possible suppression of the index $N$).

We first observe that there is a doubling rule as $N$ increases. Suppose $\u N\lambda$ is the (true) eigenvector of $g_N$ of eigenvalue $\lambda$\@. Then it is easy to show using \Eqref{eq:recursion} that
\be
g_\Np\left(
\begin{array}{c}
\u N\lambda\\
\noalign{\vskip 3pt}
\u N\lambda
\end{array}
\right)
=\lambda
\left(
\begin{array}{c}
\u N\lambda\\
\noalign{\vskip 3pt}
\u N\lambda
\end{array}
\right).
\label{eq:gdoubling}
\ee
\labeld{eq:gdoubling}
In other words, the doubled true eigenvector is a true eigenvector of the next larger (by a factor 2) ``$g$.''

\refstepcounter{remark} \label{rem:differentdoubling}
\smallskip\noindent\textsf{Remark \arabic{remark}}:~
\labeld{rem:differentdoubling}
For $w$ the doubling is slightly different. There is a minus sign and a shift in eigenvalue. One can immediately verify that if $w_N \v N \lambda=\lambda \v N \lambda$, then
\be
w_{N+1}\lsmatrixcurved{r}{\v N \lambda\\ -\v N \lambda}=(\lambda-1)\lsmatrixcurved{r}{\v N \lambda\\ -\v N \lambda}
\,.
\label{eq:wdoubling}
\ee
\labeld{eq:wdoubling}

But the doubling is more general and applies to the entire invariant space. Let $\uu N\lambda \ell$ be an element of the invariant space of $\lambda$ (for the matrix $g$), with the convention that $\ell=1$ is the true eigenvector, so that $1\leq\ell\leq N_\lambda$, with $N_\lambda$ the dimension of the invariant space (which is $\cc N{|\lambda|}$ for $\delta=0$). The others are labeled according to their place in the Jordan form in the following way:
\be
\left(g_N - \lambda I_N\right) \uu N\lambda \ell  =\uu N \lambda {\ell-1} \,,\quad\hbox{for~}2\leq\ell\leq N_\lambda\,.
\label{eq:jordanlowering}
\ee
\labeld{eq:jordanlowering}
Note that in writing $\uu N\lambda \ell$ we will often replace $\lambda$ by its negative, since this is unambiguous and allows the integer structure to stand out. Now consider $g_\Np$ applied to the doubled vector (for $\ell\geq2$),
\bea
\left(g_{N+1}-\lambda I_\Np\right)
           \left(\lsarray{c}{ \uu N\lambda \ell \\
                             \noalign{\smallskip}
                              \uu N \lambda \ell }\right)
&=&
\nonumber\\
&&\mskip -170mu
=\lsmatrix{cc}{g_N-\tilde\Delta -\lambda I_N& \tilde\Delta \\
           \noalign{\smallskip}
          \Delta & g_N-\Delta -\lambda I_N}
    \left(\lsarray{c}
            { \uu N \lambda \ell    \\
         \noalign{\smallskip}
            {\uu N \lambda \ell}  }
            \right)
\nonumber\\
&&\mskip -170mu
=\lsmatrix{cc}{g_N -\lambda I_N& 0_N \\
         \noalign{\smallskip}
        0_N & g_N -\lambda I_N}
     \left(\lsarray{c}{ \uu N\lambda \ell  \\
           \noalign{\smallskip}
          \uu N\lambda \ell }\right)
\nonumber \\
&&\mskip -170mu
=\left(\lsarray{c}{ \uu N\lambda {\ell-1}   \\
                      \noalign{\smallskip}
                     \uu N \lambda{\ell-1}
                      }\right)
\label{eq:doubling}
\,.\eea
\labeld{eq:doubling}
The final result is still a doubled vector, as a result of which the $\Delta$ and $\tilde\Delta$ entries in $g_\Np$ have no effect on it. Therefore one can continue to apply $\left(g_{N+1}-\lambda I_\Np\right)$, each time reducing $\ell$, until, reaching 1, the vector is annihilated. Since this vector was in the invariant space for $N$ it will be annihilated in at most $\cc N{|\lambda|}$ steps and therefore is certainly in the invariant space for~$\Np$\@. Physically, the feature arises because the addition of another spin does not modify the transitions between spins that are \textit{above} it \cite{luckmehta1}. As one goes from $N$ to $N+1$ spins, the spin \textit{states} are doubled; the invariant space for a particular eigenvalue of the $N$-spin configuration is also doubled to reflect this in the $N+1$-spin transition matrix, but as all the corresponding transitions are unaffected by the presence of spin $N+1$, it is left otherwise untouched.

On a lighter note, the doubling property means that for each $N+1$, $2^N$ dimensions (out of $2^{N+1}$) are accounted for, so you might say that half the job is done. Unfortunately there are many interesting questions in the other half.

\refstepcounter{remark} \label{rem:appendixoperators}
\smallskip\noindent\textsf{Remark \arabic{remark}}:~
\labeld{rem:appendixoperators}
The operations and relations discussed here can be conveniently phrased in terms of the operators defined in App.~\ref{sec:quasisymmetry}.

As observed earlier (Remarks \ref{rem:y}, \ref{rem:isospectralone} and \ref{rem:isospectraltwo}), the recursion, \Eqref{eq:recursion}, establishes the spectrum and multiplicity of the invariant space, but it does not indicate the number of true eigenvectors in each invariant space. For example, in the symmetric case ($\Delta= \one_N/2$) there is a full complement of eigenvectors. We have found however, that all invariant spaces arising from the zero-temperature granular dynamics model have but a single eigenvector (also for $\delta\neq0$, $\epsilon$ generic). We know this numerically up to $N=6$\@. We next show analytically that for our model the invariant spaces do contain non-eigenvectors (a statement that is much weaker than our numerical experience).

For $N=2$,  $\lambda=-1$, one obtains from the Jordan form representation that $(\uu211)^\top=[1,-1,1,-1]$ and $(\uu212)^\top=[2,0,0,-2]$\@. It follows that $\uu211=\left(w_2^\top+1\right) \uu112$ and that $\uu212$ is not an eigenvector. Now we look at the $\lambda=-1$ subspace for $N=3$ and study the doubling of $\uu212$: this corresponds to the separate action of $w_2^\top$ on each portion. This does not produce zero, but instead produces the $N=3$ eigenvector for  eigenvalue $-1$; i.e., the doubled $\uu212$ is in the invariant space, but is not an eigenvector. This argument obviously continues to hold for higher $N$\@. Note that the doubled objects we have produced may not coincide with what appears when obtaining the Jordan form. This is because arbitrary pieces of (e.g.) the eigenvector can be added without changing the lowering property (as exemplified in~\Eqref{eq:jordanlowering}).

\subsection{Largest magnitude eigenvalues\label{sec:largeeig}\labeld{sec:largeeig}}

The simplest eigenvector of $g$ is that with eigenvalue 0, namely the vector whose entries are all 1. There is corresponding explicitness for the \textit{largest} magnitude eigenvector, but this time we look at the eigenvector of $w$, not its transpose. Recall that for $w_N$ the largest magnitude eigenvalue is $\lambda=-N$\@.

Suppose then that
\be
w_{N+1}\left(\begin{array}{c}\alpha\\ \beta\end{array}\right)
=\lambda \left(\begin{array}{c}\alpha\\ \beta\end{array}\right)
\,,
\ee
with $\alpha$ and $\beta$ column vectors of length $2^N$\@. Then from the recursion, \Eqref{eq:recursion}, it follows that
\bea
w_N\alpha - \Delta \alpha +\tilde\Delta\beta&=&\lambda\alpha \\
w_N\beta - \tilde\Delta \beta +\Delta\alpha&=&\lambda\beta
\,.
\eea
Adding these one gets
\be
w_N\left(\alpha +\beta\right)=\lambda\left(\alpha +\beta\right)
\,.
\label{eq:wNalphaplusbeta}
\ee
Now suppose $\lambda=-N-1$\@. We know that $-(N+1)$ is \textit{not} in the spectrum of $w_N$, implying that $\beta=-\alpha$\@. Now for $w_1$ of Approach \ref{item:startinduction2} (of Sec.\ \ref{sec:recursion}) the eigenvector with eigenvalue $-1$ is
\be
\left(\begin{array}{r}\mu\\ -1\end{array}\right)
\,,
\ee
irrespective of the value of $a$ in that matrix (our case is $a=-1/2)$\@. It follows by induction that for the largest absolute value eigenvalue of $w_N$ the eigenvector consists entirely of $\pm\mu$'s  and $\pm1$'s. The pattern is built by successive attachments. For $\mu=1$ (which is the value for zero-temperature granular dynamics) the entry is $(-1)^{\nu_1}$ with $\nu_1$, the number of 1's in the state. For $\mu\neq1$, the odd entries are multiplied by $\mu$\@. If one uses spin notation, i.e., $\mu_\ell\to\sigma_\ell=2\mu_\ell-1$, then this largest magnitude eigenvector is given by the product of the spins, $\sigma_1 \sigma_2 \ldots \sigma_N$\@.

The simplicity of this eigenvector is to be compared to the corresponding eigenvector of $g_N$\@. First note that by appropriate multiplication by powers of 2, all vectors in the invariant spaces can be written as integers. We have observed that (with minimal multiplication) the elements of these vectors grow rapidly. For example, the first component of $\uu 661$ is 2075566815213212256361\@. Factoring this number (it equals $19\! \times\! 41 \!\times \!193 \!\times\! 239\! \times\! 18523\! \times\! 3118406479$) or others like it has not given us clues to its structure. The algorithm for obtaining eigenvectors involves sums and products of the matrix elements, so it would appear that the many halves in $g_N$ are not canceling in any systematic way. It should be noted that this growth applies not only to $\uu NN1(1)$, but to pretty much any component for which there was no neat explicit form.

\refstepcounter{remark} \label{rem:six}
\smallskip\noindent\textsf{Remark \arabic{remark}}:~
\labeld{rem:six}
Consider, for $w_{N+1}$, the eigenvalue $N$, which \textit{is} an eigenvalue of $w_N$ as well. Then \Eqref{eq:wNalphaplusbeta} implies that $\alpha+\beta$ is an eigenvector of $w_N$\@. Assuming that $\alpha+\beta$ is not zero, then $\beta=\uu NN1-\alpha$, with $\uu NN1$ a true eigenvector of $w_N$\@.

\subsection{On the ``maximal Jordan'' property\label{sec:maxjordan}\labeld{sec:maxjordan}}

All zero-temperature granular models give rise either to what we have called ``maximally Jordan'' spectrum or to complex spectrum. Thus all $\eps$ and $\delta$ except those combinations yielding fish bone spectrum (i.e,. generic $\eps$ or non-generic $\eps$, zero $\delta$), are maximally Jordan. (Recall, this means that degenerate eigenvalues, of which we assume at least one present, have but a single eigenvector. Recall too that $|\delta|<1/2$.) This assertion is based on numerical solutions up to and including~$N=6$\@. Included in this observation are the smaller subspaces for $\eps$ generic, $\delta$ nonzero.

On the other hand, if we use the recursion of \Eqref{eq:recursion} and generate $\Delta$'s whose diagonal entries are 0's $1/2$'s and 1's (but which need not be realizations of the granular model), sometimes one gets the maximal Jordan property, sometimes not. In this subsection we explore further criteria for this property.

Consider the eigenvalue equation for $w_{N+1}$ ($N+1$  spins)
\be
w_{N+1}\xi=\lsmatrix{cc}
              {w_N-\widetilde\Delta_N   &     \Delta_N\\
               \widetilde\Delta_N       &  w_N-\Delta_N}
               \lsmatrixcurved{c}
               {x\\ y}
=              \lsmatrixcurved{c}
               {w_Nx-x+\Delta_N(x+y)\\ w_Ny-y+\widetilde\Delta_N(x+y)}
=           -k    \lsmatrixcurved{c}
                  {x\\ y}
\,,
\label{eq:0eig}
\ee
\labeld{eq:0eig}
with $x$ and $y$ $2^N$-vectors, $k=0,1,\dots,N$ and use has been made of $\Delta+\widetilde\Delta=\one$\@. We only consider $\delta=0$\@. Add these equations to get
\be
w_N(x+y)=-k(x+y)
\label{eq:0sum}
\,.
\ee
\labeld{eq:0eig}
We make the inductive hypothesis that \textit{every invariant space of $w$ has but a single eigenvector}, which in the presence of degenerate eigenvalues (which we know occur for $N>1$) is the maximal Jordan condition. Then \Eqref{eq:0sum} implies that $x+y=\alpha v_k$, with $v_k$ \textit{the} eigenvector of $w_N$ of eigenvalue $-k$\@. $\alpha$ is a real number and there are two possibilities: it is zero or it is not. If it's not, renormalize $x$ and $y$ to make it $-1$\@.\\
Case I. $\alpha=0$\@. In this case $x=-y$\@. From \Eqref{eq:0eig} we have
\be
w_N x=-(k-1)x
\,,
\ee
so that $x$ is an eigenvector with eigenvalue $-k+1$ and we have (re-) discovered the doubling property, namely that once you have a true eigenvector of eigenvalue $-k$ you can get the (or ``a'' if it's not maximally Jordan) eigenvector (of eigenvalue $-(k-1)$) for one more spin by doubling and changing sign.\\
Case II. $\alpha=-1$\@. Then $x+y=-v_k$ and we use \Eqref{eq:0sum} to obtain
\bea
w_N x +(k-1)x&=&\Delta_N v_k  \nonumber\\
w_N y +(k-1)y&=&\widetilde\Delta_N v_k
\label{eq:nonzeroalpha}
\eea
\labeld{eq:nonzeroalpha}
To prove the maximal Jordan property we need to show that \Eqref{eq:nonzeroalpha} has no solution. In that way, moving to higher $N$, you never get a new eigenvector, just doubling of the old ones---with one exception. That exception is $k=0$ and that's the new eigenvector (of $w$, not $w^\top$).

\refstepcounter{remark} \label{rem:0alpha}
\smallskip\noindent\textsf{Remark \arabic{remark}}:~
\labeld{rem:0alpha}
For $k=0$, \Eqref{eq:nonzeroalpha} can always be solved, since $(+1)$ is not an eigenvalue of $w$, and the left hand side is invertible. As remarked elsewhere in this article, the $k=0$ eigenvector of $w$, when scaled to involve integers with no common denominator, has entries that grow faster than exponentially (and to us, unpredictably) with~$N$\@. These entries, normalized to sum to unity, are the probabilities of the various spin configurations in the stationary state. As indicated below, the nature of the stationary state has a sensitive dependence on the parameter $\eps$ and on the boundary conditions.

Considering now only the $k>0$ case, the left hand side of \Eqref{eq:nonzeroalpha} has an important property: it cannot produce a vector proportional to $v_{k-1,n}$, where $n=C^N_{k-1}$, and we use the notation (of Sec.\ \ref{sec:eigenvectors}) indicating that $v_{k-1,n}$ is the deepest vector in the Jordan chain. This vector is characterized by the property that it, and only it, can survive $n-1$ applications of $\left(w+(k-1)\one\right)$\@. (It is not unique, but can be selected uniquely by the demand that it be orthogonal to the rest of its invariant space. But even without selection the foregoing criterion obtains.)

So \textit{the problem of showing this system to be maximally Jordan reduces to showing that either $\Delta v_k$ or $\widetilde\Delta v_k$ has non zero overlap with $v_{k-1,n}$ (with $n=C^N_{k-1}$)}, the deepest vector in the Jordan chain.\@.

\refstepcounter{remark} \label{rem:necsuff}
\smallskip\noindent\textsf{Remark \arabic{remark}}:~
\labeld{rem:necsuff}
This condition is necessary \textit{and} sufficient. This means that all the maximally Jordan matrices generated with $\Delta$'s that do not correspond to what you'd get from the granular model also have no solution to the above equation. So the criterion is more general than the model, but for dynamical reasons the model satisfies it with each additional spin.

\refstepcounter{remark} \label{rem:lowestJordan}
\smallskip\noindent\textsf{Remark \arabic{remark}}:~
\labeld{rem:lowestJordan}
Like the stationary state, this lowest vector in the Jordan chain is created \textit{de novo} with each succeeding increase in $N$, and is not obtained by doubling. This is clear from \Eqref{eq:doubling}, since vectors in the invariant space obtained by doubling cannot require the maximal number of steps for the increased $N$ to reach the eigenstate under applications of $(w+k\one)$\@.

\refstepcounter{remark} \label{rem:orthog}
\smallskip\noindent\textsf{Remark \arabic{remark}}:~
\labeld{rem:orthog}
A word of caution regarding the establishing of whether or not  $\widetilde\Delta x$ or $\Delta y$ have components along some direction in the vector space: as usual when the Jordan form is needed, many of the usual tools break down. Thus even for a general stochastic matrix, if it does not require a Jordan form you can still have $\mathrm{left}_\alpha^\dagger \cdot \mathrm{right}_\beta=\delta_{\alpha\beta}$, where ``left'' and ``right'' are eigenfunctions of $w$ and $w^T$ transpose respectively (so that ``left$^\dagger$'' is a left eigenvector of $w$). When $w$ requires a Jordan form, this is not true and for our maximally Jordan matrices the left and right eigenvectors (when the invariant space is of dimension greater than one) are orthogonal.

\refstepcounter{remark} \label{rem:gmax}
\smallskip\noindent\textsf{Remark \arabic{remark}}:~
\labeld{rem:gmax}
Because the doubling properties of $w$ and its transpose (``$g$'') differ slightly, the corresponding equations for $g$ take the following form
\bea
g_N x +k x &=&~~\widetilde\Delta_N u_{k-1}  \nonumber\\
g_N y +ky&=&-\Delta_N u_{k-1}
\label{eq:nonzeroalphag}
\,.
\eea
\labeld{eq:nonzeroalphag}
The condition for proving the maximal Jordan property is that application of $\Delta$ or $\widetilde\Delta$ to the next slower eigenvector (i.e., $-\lambda=k-1$) not yield any component along the bottom of the Jordan chain for eigenvalue $-k$\@. This condition is particularly simple for $k=1$ (since $u_0$ is all 1's), and we have noted numerically that the lowest in the $k=1$ Jordan chain is indeed highly correlated with the associated $\Delta$\@.

\section{Time dependence \label{sec:timedependence}\labeld{sec:timedependence}}

Non-monotonic time evolution occurs in this model, just as for a critically damped oscillator. From the definition of $w$, the time dependence of an $N$-spin system is
\be
p(t)=\exp(w_Nt)p(0)
\,,
\ee
with $p$ the (vector) probability distribution (at times 0 and $t$). Were this a situation with (only) bona fide eigenvectors one would expect $p(t)$ to be a sum of terms of the form $\exp(\lambda_\alpha t)$ with $\alpha$ running over the eigenvalues. With eigenvalue degeneracy, terms of the form $t^k\exp(\lambda_\alpha t)$ can occur, for $k$ up to one less than the level of degeneracy. In this section we show precisely how the power law enters in the probability distribution. In Sec.\ \ref{sec:dyn2s} it will emerge naturally in a study of correlations.

\begin{figure}
\includegraphics[angle=270, width=0.6\textwidth]{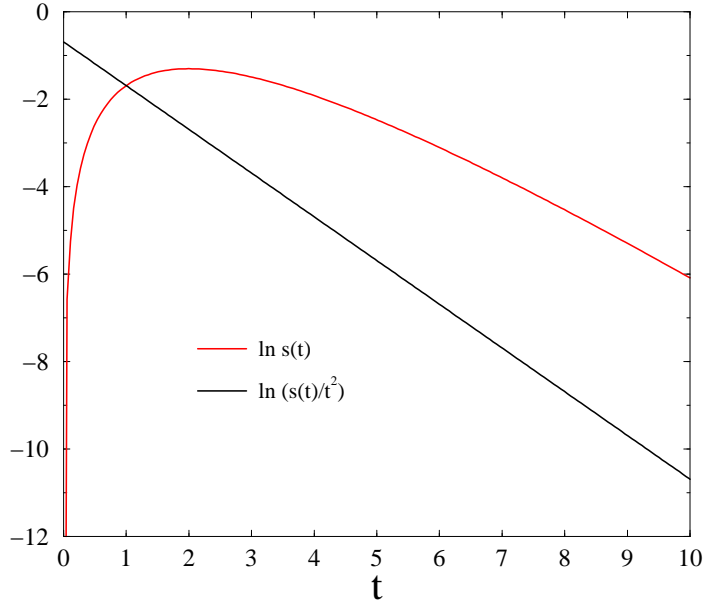}
\caption{Survival amplitude, $s(t)$, of an initial state of the form $p_0+\rho\vv 312$\@. The stationary state $p_0$ is removed and an appropriate projection of the remainder is taken. What is shown is the logarithm of what's left, both in raw form and after division by $t^2$\@. Note that this is the 3-spin generating matrix and we are looking at the invariant space associated with the eigenvalue -1.
\label{fig:powerlaw}\labeld{fig:powerlaw}}
\end{figure}

Let the initial state be of the form $p_0+\rho\vv N\ell\ell$ with $p_0$ the stationary distribution (a.k.a.\ $\vv N01$) and $\rho$ small enough so that all components of $p(0)$ are non-negative. The action of $w$ on this is as follows: it annihilates $p_0$; the $\vv N\ell\ell$ term is also annihilated but two new terms replace it: one is proportional to the eigenvector ($\vv N\ell{1}$) and the other is $\vv N\ell{\ell-1}$\@. The exponentiation of $w$ is not quite so simple as when there is a spectral decomposition, but it remains true that this process respects the invariant spaces. Moreover, one can immediately verify that for
\be m=\left(
       \begin{array}{ccccc}
       \lambda & 1       & 0  & 0        & \dots\\
       0       & \lambda & 1  & 0        &\dots \\
       \vdots  & \dots   & \ddots        & \vdots       \\
       0       & \dots   &    & \lambda  &   1   \\
       0       & \dots   &    &   0      &  \lambda
       \end{array}
       \right)
\label{eq:upperdiag}
\ee
the matrix exponential is
\be \exp(tm)=\e^{t\lambda}\left(
       \begin{array}{cccccccc}
       1       & t       & t^2/2     & t^3/6   &~&~& \dots   & t^{\ell-1}/(\ell-1)!\\
       0       & 1       & t         & t^2/2   & & &         &\dots    \\
       \vdots  & \dots   & \ddots    & \vdots  & & &          \\
       0       & \dots   &           &         & & &   1     &  t   \\
       0       & \dots   &           &         & & &   0     &  1
       \end{array}
       \right)
\,.
\ee
That is, $\e^{-t\lambda}\exp(tm)$ is upper triangular with the $k^\mathrm{th}$ ascending diagonal constant and equal to $t^k/k!$, $k=0,\dots,\ell-1$, with $\ell$ the dimension of $m$\@. To see the power law for the initial state given above, we require some notation. Let the Jordan decomposition of $w$ be accomplished by the following transformation
\be
wV=Vj
\,,
\ee
with $j$ composed of blocks of the form given in \Eqref{eq:upperdiag}. Let $u^\top$ be the (row) vector satisfying $u^\top\vv N\ell1=1$ and which is orthogonal to all other vectors (it is given by an appropriately normalized row of $V^{-1}$). With this notation, the object having power law decay is $s(t)\equiv u^\top\exp(wt)\vv N\ell\ell$\@. This is illustrated in Fig~\ref{fig:powerlaw}.

\section{Dynamical equations for spin correlations\label{sec:spincorrelations}\labeld{sec:spincorrelations}}

\subsection{Background}

In this section we investigate the dynamics of the model based on (equal-time) spin correlations. These quantities are defined as the time-dependent mean values of products of spins:
\begin{equation}
M_{i_1\dots i_k}(t)=\mean{\s_{i_1}(t)\cdots\s_{i_k}(t)},
\end{equation}
where $i_1,\dots,i_k$ is an ordered $k$-tuple of distinct labels in $1,\dots,N$\@. It is sufficient to consider multilinear functions, i.e., functions which are at most linear in each of the spins, since each spin obeys $\s_i^2=1$\@. There are $\cc{N}{k}$ such correlations at level $k$, and hence $2^N$ correlations in total (including $M_\emptyset=1$). The spin correlations thus form a basis of observables. The knowledge of those correlations is tantamount to that of all configuration probabilities.

The temporal evolution of spin correlations is dictated by first-order linear differential equations, which can be written down explicitly.

To illustrate the approach, we first consider the infinite-temperature situation, i.e., the random walk on the hypercube in dimension $N$, mentioned in Remark \ref{rem:y}. In this case, each spin flips at Poissonian times with unit rate, independent of the others. Consider for definiteness the evolution of the first spin $\s_1(t)$ during an infinitesimal time interval $\d t$\@. We have
\begin{equation}
\s_1(t+\d t)=\left\{
\begin{matrix}
\hfill\s_1(t)&&\hbox{with prob.}&1-\d t,&\\
-\s_1(t)&&\hbox{with prob.}&\d t.&
\end{matrix}
\right.
\end{equation}
The average, $M_1(t)=\mean{\s_1}$, therefore obeys $M_1(t+\d t)=(1-2\d t)M_1(t)$\@. As a consequence
\begin{equation}
\der{M_1}=-2M_1,
\end{equation}
and more generally
\begin{equation}
\der{M_{i_1\dots i_k}}=-2kM_{i_1\dots i_k}.
\label{eq:corrdiffeq}
\end{equation}
\labeld{eq:corrdiffeq}
The basis of spin correlations therefore diagonalizes the dynamics. We thus readily obtain the spectrum of the Markov matrix. The eigenvalue $\lambda=-2k$ has multiplicity $\cc{N}{k}$ (in accordance with the number of possible subsets associated with \Eqref{eq:corrdiffeq}), and the quantities $M_{i_1\dots i_k}$ provide an explicit set of eigenvectors for the infinite temperature model. This diagonalization is equivalent to that described in Remark~\ref{rem:y}.

\subsection{General approach}

Let us apply the above approach to our column model.

\begin{itemize}

\item
If $h_n\ne0$, $\s_n$ is updated with unit rate according to the rule
\begin{equation}
\s_n\to\sign h_n.
\label{eq:update1}
\end{equation}
\labeld{eq:update1}

\item
If $h_n=0$, $\s_n$ is updated with rate $w=\half+\delta$ \cite{note:wnotation} according to the rule
\begin{equation}
\s_n\to-\s_n.
\label{eq:update2}
\end{equation}
\labeld{eq:update2}

\end{itemize}

For definiteness, we start with the following three hypotheses:

\begin{itemize}

\item[(1)]
Free boundary conditions (i.e., $h_1=0$).

\item[(2)]
$w=\half$ (i.e., $\delta=0$).

\item[(3)]
$\eps$ is generic (i.e., irrational or `large' rational).

\end{itemize}

As discussed earlier, the phrase ``$\eps$ is not generic'' means that $\eps$ is one of the special values that allows vanishing $h$ below the first spin (see Table~\ref{table:list}), i.e., for given system size, $N$, $\eps=p/q$ (irreducible) with $p+q\le N-1$\@.

For $w=\half$, the rule~(\ref{eq:update1}) still holds {\it on average} for $h_n=0$ (defining $\sign 0\equiv0$). Flipping spin $\s_n$ with probability $\half$ sets $\s_n=\pm1$ with equal probability. Therefore in full generality
\begin{equation}
\der{M_n}=-M_n+\mean{\sign h_n}\,,
\end{equation}
where, as above, $M_n=\langle\sigma_n\rangle$\@. The same observation will allow us to write down differential equations for all the observables, as they are multilinear functions, i.e., linear functions of each of the spins.

The above approach is to be put in perspective with the pioneering work of Glauber on the dynamics of the ferromagnetic Ising chain~\cite{glauber}. Although Glauber only considered one-spin correlations (magnetization profile) and two-spin correlations in his original work, several subsequent papers have been devoted to a systematic extension to an arbitrary number of spins~\cite{bedeaux, felderhof,aliev, mayer}.

Before considering the general situation, it is useful to first derive explicit equations for the first few values of $N$\@.

\subsection{The case $N=1$}

There is only one non-trivial correlation, $M_1(t)=\mean{\s_1(t)}$, which obeys the differential equation
\begin{equation}
\der{M_1}=-M_1+\mean{\sign h_1},
\end{equation}
with
\begin{equation}
\sign h_1=0,
\label{eq:sign1}
\end{equation}
\labeld{eq:sign1}
so that
\begin{equation}
\der{M_1}=-M_1.
\label{eq:m1dot}
\end{equation}
\labeld{eq:m1dot}
This equation for $M_1$ still holds for larger system sizes, because the motion of spin-1 does not depend on those below it. We thus obtain
\begin{equation}
M_1(t)=M_1(0)\mskip1mu\e^{-t}.
\end{equation}

\subsection{The case $N=2$}

Besides $M_1(t)$, there are two new correlation functions at size $N=2$, $M_2(t)=\mean{\s_2(t)}$ and
$M_{12}(t)=\mean{\s_1(t)\s_2(t)}$\@. They obey the differential equations
\begin{eqnarray}
\der{M_2}&=&-M_2+\mean{\sign h_2},\nonumber\\
\der{M_{12}}&=&-2M_{12}+\mean{\sign h_1\cdot\s_2}+\mean{\s_1\cdot\sign h_2},
\end{eqnarray}
with
\begin{equation}
\sign h_2=-\s_1\,.
\label{eq:sign2}
\end{equation}
\labeld{eq:sign2}
Note that this holds irrespective of $\eps$\@. It follows that
\begin{eqnarray}
\der{M_2}&=&-M_2-M_1,\nonumber\\
\der{M_{12}}&=&-2M_{12}-1.
\label{eq:m2dot}
\end{eqnarray}
\labeld{eq:m2dot}
These equations continue to hold for larger system sizes, because of the dynamics of a given spin is independent of those below it.

A detailed analysis of the dynamics of the model with $N=2$, albeit with a general $w=\half+\delta$, will be performed in Section~\ref{sec:dyn2s}.

\subsection{The case $N=3$}

There are 4 new correlation functions at size $N=3$, which obey the differential equation
\begin{eqnarray}
\der{M_3}&=&-M_3+\mean{\sign h_3},\nonumber\\
\der{M_{13}}&=&-2M_{13}+\mean{\sign h_1\cdot\s_3}+\mean{\s_1\cdot\sign h_3},\nonumber\\
\der{M_{23}}&=&-2M_{23}+\mean{\sign h_2\cdot\s_3}+\mean{\s_2\cdot\sign h_3},\nonumber\\
\der{M_{123}}&=&-3M_{123}+\mean{\sign h_1\cdot\s_2\s_3}+\mean{\s_1\cdot\sign h_2\cdot\s_3}+\mean{\s_1\s_2\cdot\sign h_3},
\end{eqnarray}
with
\begin{equation}
\sign h_3=\half(\s_1\s_2-\s_1-\s_2-1).
\label{eq:sign3}
\end{equation}
\labeld{eq:sign3}
This identity holds whenever $\eps<1$\@. Assuming this for definiteness, we have
\begin{eqnarray}
\der{M_3}&=&-M_3+\half(M_{12}-M_1-M_2-1),\nonumber\\
\der{M_{13}}&=&-2M_{13}+\half(-M_{12}-M_1+M_2-1),\nonumber\\
\der{M_{23}}&=&-2M_{23}-M_{13}+\half(-M_{12}+M_1-M_2-1),\nonumber\\
\der{M_{123}}&=&-3M_{123}-M_3+\half(-M_{12}-M_1-M_2+1).
\label{eq:m3dot}
\end{eqnarray}
\labeld{eq:m3dot}

\subsection{The general situation\label{sec:general}\labeld{sec:general}}

The structure of the dynamical equations is apparent from the above examples. The general procedure consists of the following steps.

A general spin correlation $M_{i_1\dots i_k}$ obeys a differential equation of the form
\begin{equation}
\der{M_{i_1\dots i_k}}=-kM_{i_1\dots i_k}+\sum_{j=1}^k H^{(j)}_{i_1\dots i_k},
\label{eq:mder}
\end{equation}
\labeld{eq:mder}
where
\begin{equation}
H^{(j)}_{i_1\dots i_k}=\mean{\s_{i_1}\dots\s_{i_{j-1}}\cdot\sign h_{i_j}
\cdot\s_{i_{j+1}}\dots\s_{i_k}}
\label{eq:hdef}
\end{equation}
\labeld{eq:hdef}
is obtained from $M_{i_1\dots i_k}$ by replacing the $j^\mathrm{th}$ spin $\s_{i_j}$
by $\sign h_{i_j}$\@.

For each depth $n$, $\sign h_n$ is a symmetric multilinear function of the spins above $n$, i.e., $\s_1,\dots,\s_{n-1}$\@. Its explicit expression can be obtained by listing the values of $\sign h_n$ for all the $2^{n-1}$ spin configurations, and fitting this set of values to a multilinear function with arbitrary coefficients. The resulting expression, generalizing~(\ref{eq:sign2}) and~(\ref{eq:sign3}), depends on $\eps$\@. More precisely, it depends on the relative position of $\eps$ with respect to the special values that enter at depth $n$, such that $h_n$ can vanish. These are the $(n-2)$ rationals $\eps=p/q$ (not necessary irreducible) such that $p+q=n-1$\@. The expression of $\sign h_n$ is constant for $\eps$ between any two consecutive special values. Table~\ref{table:list} gives an ordered list of the special values of $\eps$ corresponding to the first few values of $n$\@.

\begin{table}[!ht]
\begin{center}
\begin{tabular}{|c|c|}
\hline
\ $n$ \ &\ special values of $\eps$\ \\
\hline
3&1\\
\hline
4&$\frac12$, 2\\
\hline
5&$\frac13$, 1, 3\\
\hline
6&$\frac14$, $\frac23$, $\frac32$, 4\\
\hline
7&$\frac15$, $\frac12$, 1, 2, 5\\
\hline
8&$\frac16$, $\frac25$, $\frac34$, $\frac43$, $\frac52$, 6\\
\hline
9&$\frac17$, $\frac13$, $\frac35$, 1, $\frac53$, 3, 7\\
\hline
\end{tabular}
\end{center}
\caption{Special values of $\eps$ when the total number of spins is $n$\@.}
\label{table:list}\labeld{table:list}
\end{table}

Now, for a fixed generic $\eps$, let us replace in~(\ref{eq:hdef}) $\sign h_{i_j}$ by its explicit expression in terms of the spins, carry this into~(\ref{eq:mder}), and expand the sum. The expression thus obtained is a linear combination of operators whose labels are strictly smaller than that of $M_{i_1\dots i_k}$\@. The label $\ell$ of an operator (spin product) $\s_{i_1}\dots\s_{i_k}$ is defined as
\begin{equation}
\ell=2^{i_1-1}+2^{i_2-1}+\cdots+2^{i_k-1}.
\end{equation}
This labeling provides a universal ordering of operators for all system sizes, whose beginning is given in Table~\ref{table:labels}. The ordering of $\ell=11$ and $\ell=12$ is the first surprising or unexpected one, in the sense that the number of spins involved decreases at fixed $N$\@.

\begin{table}[!ht]
\begin{center}
\begin{tabular}{|c|c|c|}
\hline
$N$&$\ell$&operator\\
\hline
0&0&1\\
\hline
1&1&$\s_1$\\
\hline
2&2&$\s_2$\\
&3&$\s_1\s_2$\\
\hline
3&4&$\s_3$\\
&5&$\s_1\s_3$\\
&6&$\s_2\s_3$\\
&7&$\s_1\s_2\s_3$\\
\hline
4&8&$\s_4$\\
&9&$\s_1\s_4$\\
&10&$\s_2\s_4$\\
&11&$\s_1\s_2\s_4$\\
&12&$\s_3\s_4$\\
&13&$\s_1\s_3\s_4$\\
&14&$\s_2\s_3\s_4$\\
&15&\ $\s_1\s_2\s_3\s_4$\ \\
\hline
\end{tabular}
{\hskip 15pt}
\begin{tabular}{|c|c|c|}
\hline
$N$&$\ell$&operator\\
\hline
5&16&$\s_5$\\
&17&$\s_1\s_5$\\
&18&$\s_2\s_5$\\
&19&$\s_1\s_2\s_5$\\
&20&$\s_3\s_5$\\
&21&$\s_1\s_3\s_5$\\
&22&$\s_2\s_3\s_5$\\
&23&$\s_1\s_2\s_3\s_5$\\
&24&$\s_4\s_5$\\
&25&$\s_1\s_4\s_5$\\
&26&$\s_2\s_4\s_5$\\
&27&$\s_1\s_2\s_4\s_5$\\
&28&$\s_3\s_4\s_5$\\
&29&$\s_1\s_3\s_4\s_5$\\
&30&$\s_2\s_3\s_4\s_5$\\
&31&\ $\s_1\s_2\s_3\s_4\s_5$\ \\
\hline
\end{tabular}
\end{center}
\caption{Universal labeling of operators.}
\label{table:labels}\labeld{table:labels}
\end{table}

The dynamical equations~(\ref{eq:mder}) therefore have a triangular form, for any finite system size $N$\@. The explicit form of those equations depends on $\eps$, albeit their triangular structure is robust. This feature provides an alternative proof of the pattern of degeneracies of the Markov matrix studied earlier by algebraic techniques. We indeed obtain at once from the triangular form of~(\ref{eq:mder}) that the spectrum consists of the negative integers, i.e., $\lambda=-k$ ($k=0,\dots,N$), with combinatorial multiplicities
$\cc{N}{k}$\@. Furthermore, although the occurrence of a single Jordan block of maximal length $\cc{N}{k}$ at each level $k$ is not proved by this approach, this generic feature is suggested by the large number of off-diagonal terms generated by expanding the sum in the right-hand side of~(\ref{eq:mder}).

\subsection{Extensions\label{sec:extensions}\labeld{sec:extensions}}

We extend the above approach by investigating whether and how the three hypotheses made below~(\ref{eq:update2}) can be lifted.

\begin{itemize}

\item
Lifting hypothesis (1) is easy. Let us remark that hypothesis (1) corresponds to having free boundary conditions. Indeed the field $h_1$ vanishes identically, so that the uppermost spin $\s_1$ is free. A natural alternative consists of fixing the uppermost spin, along the lines of our previous works~\cite{mehtaluck4, luckmehta1, luckmehta2, luckmehta3}. Let us rename this spin $\s_0$ and set $\s_0=+1$ for definiteness. This amounts to adding a constant to all the fields $(h_n\to h_n+f(\s_0)=h_n+f(+1)=h_n-1)$\@. Accordingly, the system size (number of degrees of freedom) is reduced from $N$ to $N-1$\@. The explicit multilinear expressions of $\sign h_n$ in terms of the spins are therefore modified. The overall construction however still holds. The spectrum is left unchanged (up to the size reduction from $N$ to $N-1$).

\item
Lifting hypothesis (2) ($w=\half$) is also easy, provided hypothesis (3) is maintained. Indeed, as long as $\eps$ is irrational, the only point of zero field where $h_n$ can vanish is $n=1$\@. As a consequence, the zero-field rate $w=\half+\delta$ only affects the dynamics of the uppermost spin $\s_1$\@. Therefore, it only enters the diagonal term of the differential equation~(\ref{eq:mder}) for the operators involving $\s_1$ (i.e., such that $i_1=1$) as follows:
\begin{eqnarray}
i_1=1&:&\qquad\der{M_{i_1\dots i_k}}
=-(k+2\delta)M_{i_1\dots i_k}+\sum_{j=1}^k H^{(j)}_{i_1\dots i_k},\nonumber\\
i_1\ne1&:&\qquad\der{M_{i_1\dots i_k}}
=-kM_{i_1\dots i_k}+\sum_{j=1}^k H^{(j)}_{i_1\dots i_k}\,.
\label{lift2}
\end{eqnarray}
\labeld{lift2}
The first line yields $\lambda=-(k+2\delta)$ with multiplicity $\cc{N-1}{k-1}$ for $k=1,\dots,N$, whereas the second one yields $\lambda=-k$ with multiplicity $\cc{N-1}{k}$ for $k=0,\dots,N-1$\@. We thus recover the known split spectrum, derived algebraically from Eq.~\parenref{eq:spectrumwithdelta}.

\item
Lifting hypothesis (3) ($\eps$ generic) is more subtle. The situation where $\eps$ is rational and $w\ne\half$ (i.e., $\delta\ne0$) is indeed exceptional, in the sense that the spectrum is not integrable. The spectrum will be shown later to generally have a complex `fish bone' structure. The gist of this complexity is as follows. For $\eps=p/q$ (irreducible), there are internal points of zero field, that is, $h_n$ can vanish. This occurs at depths $n$ such that $n-1$ is a multiple of the period $(p+q)$\@. As a consequence, for $\eps$ rational and $w\ne\half$, the total exit rate from a configuration depends on the configuration. Hence there is no Glauber rule to construct differential equations. The whole construction breaks down.

\end{itemize}

To summarize, the approach based on spin correlations works everywhere except in the non-integrable situation where $\eps$ is rational and $w\ne\half$, where there are good reasons why it cannot work.

Table~\ref{table:spectra} recapitulates our results on the spectrum of the Markov matrix, as a function of $\eps$ and $\delta=w-\half$\@. The last column gives the exact expression of the average absolute eigenvalue $\Lambda$~(see~(\ref{eq:sumdef})), calculated later as a sum rule.

\begin{table}[!ht]
\begin{center}
Free boundary conditions
\begin{tabular}{|c|c|c|c|}
\hline
$\eps$&$\delta$&spectrum&$\Lambda$\\
\hline
\ generic\ &\ $\delta=0$ \ &$\lambda=-k$\qquad mult.\ $\cc{N}{k}$\qquad($k=0,\dots,N$)&$\frac{N}{2}$\\
\hline
generic&$\delta\ne0$&
$\left\{
\begin{matrix}
\lambda=-k\hfill\\\lambda=-(k+1+2\delta)
\end{matrix}
\right.
$\qquad mult.\ $\cc{N-1}{k}$\qquad ($k=0,\dots,N-1$)&$\frac{N}{2}+\delta$\\
\hline
special&$\delta=0$&$\lambda=-k$\qquad mult.\ $\cc{N}{k}$\qquad($k=0,\dots,N$)&$\frac{N}{2}$\\
\hline
special&$\delta\ne0$&complex `fish bone'&\ $\frac{N}{2}+A_N\delta$\ \\
\hline
\end{tabular}

{\vskip 10pt}

Fixed boundary conditions

\begin{tabular}{|c|c|c|c|}
\hline
$\eps$&$\delta$&\ {\hskip 115pt}\ spectrum\ {\hskip 115pt}\ &$\Lambda$\\
\hline
\ generic\ &\ $\delta=0$ \ &$\lambda=-k$\qquad mult.\ $\cc{N}{k}$\qquad($k=0,\dots,N$)&$\frac{N}{2}$\\
\hline
generic&$\delta\ne0$&$\lambda=-k$\qquad mult.\ $\cc{N}{k}$\qquad($k=0,\dots,N$)&$\frac{N}{2}$\\
\hline
special&$\delta=0$&$\lambda=-k$\qquad mult.\ $\cc{N}{k}$\qquad($k=0,\dots,N$)&$\frac{N}{2}$\\
\hline
special&$\delta\ne0$&complex `fish bone'&\ $\frac{N}{2}+B_N\delta$\ \\
\hline
\end{tabular}
\end{center}
\caption{Spectrum and average absolute eigenvalue $\Lambda$~(see~(\ref{eq:sumdef})) of the Markov matrix of a system of size $N$, as a function of $\eps$ and $\delta=w-\half$\@. Top: free boundary conditions. Bottom: fixed boundary conditions.}
\label{table:spectra}\labeld{table:spectra}
\end{table}

\section{Full dynamics in the case of two spins \label{sec:dyn2s}\labeld{sec:dyn2s}}

In this section we give the full solution to the dynamics of the model in the first non-trivial case of two spins ($N=2$). In this situation, the correspondence between the spin correlations and the configuration probabilities $p_{\s_1\s_2}$ reads
\begin{eqnarray}
1=M_\emptyset&=&p_{++}+p_{+-}+p_{-+}+p_{--},\nonumber\\
M_1&=&p_{++}+p_{+-}-p_{-+}-p_{--},\nonumber\\
M_2&=&p_{++}-p_{+-}+p_{-+}-p_{--},\nonumber\\
M_{12}&=&p_{++}-p_{+-}-p_{-+}+p_{--},
\end{eqnarray}
i.e.,
\begin{eqnarray}
p_{++}&=&\quarter(1+M_1+M_1+M_{12}),\nonumber\\
p_{+-}&=&\quarter(1+M_1-M_1-M_{12}),\nonumber\\
p_{-+}&=&\quarter(1-M_1+M_1-M_{12}),\nonumber\\
p_{--}&=&\quarter(1-M_1-M_1+M_{12}).
\end{eqnarray}

At variance with the above general analysis, we consider an arbitrary zero-field rate $w=\half+\delta$\@. Therefore, irrespective of $\eps$, the upper spin $\s_1$ flips freely (see~(\ref{eq:sign1})) with rate $w$, whereas the lower spin $\s_2$ flips with unit rate under the action of $\sign h_2=-\s_1$ (see~(\ref{eq:sign2})). The dynamical equations for the spin correlations therefore read
\begin{eqnarray}
\der{M_1}&=&-(1+2\delta)M_1,\nonumber\\
\der{M_2}&=&-M_2-M_1,\nonumber\\
\der{M_{12}}&=&-2(1+\delta)M_{12}-1.
\label{eq:2dyn}
\end{eqnarray}
\labeld{eq:2dyn}
These dynamical equations have the expected triangular form, and yield the expected spectrum: $\lambda=0$, $\lambda=-1$, $\lambda=-(1+2\delta)$, $\lambda=-2(1+\delta)$\@.

For arbitrary initial conditions, the general solution of the above equations reads
\begin{eqnarray}
M_1(t)&=&M_1(0)\,\e^{-(1+2\delta)t},\nonumber\\
M_2(t)&=&M_2(0)\,\e^{-t}-M_1(0)R(t),\nonumber\\
M_{12}(t)&=&C_{12}+(M_{12}(0)-C_{12})\,\e^{-2(1+\delta)t},
\label{eq:2sol}
\end{eqnarray}
\labeld{eq:2sol}
where we have introduced the stationary-state correlation
\begin{equation}
C_{12}=\mean{\s_1\s_2}_{\rm stat}=-\frac{1}{2(1+\delta)}=-\frac{1}{2w+1}
\end{equation}
and the off-diagonal response function
\begin{equation}
R(t)=\frac{1-\e^{-2\delta t}}{2\delta}\,\e^{-t}.
\end{equation}

The stationary-state correlation $C_{12}$ is always negative, and it depends on the rate $w$ in a smooth way. The regimes of small and large $w$ can be understood as follows. In the $w\to0$ limit, $\s_1$ moves very slowly, and so $\s_2$ follows $\sign h_2=-\s_1$ almost perfectly adiabatically, so that $\s_2\approx-\s_1$ and $C_{12}$ tends to $-1$\@. In the opposite regime ($w\to\infty$), $\s_1$ moves so fast that $\s_2$ hardly feels a driving field, so that $C_{12}$ approaches 0.

The response function $R(t)$ is more interesting. For $\delta=0$ it reads
\begin{equation}
R(t)=t\,\e^{-t}.
\end{equation}
The resonance phenomenon recognized through the presence of the `secular' prefactor $t$ takes place precisely at the point where the eigenvalues $\lambda=-1$ and $\lambda=-(1+2\delta)$ become degenerate, so that a Jordan block of size $\cc{2}{1}=2$ is needed, in agreement with the general theory.

\begin{figure}[!ht]
\begin{center}
\includegraphics[angle=-90,width=.45\linewidth]{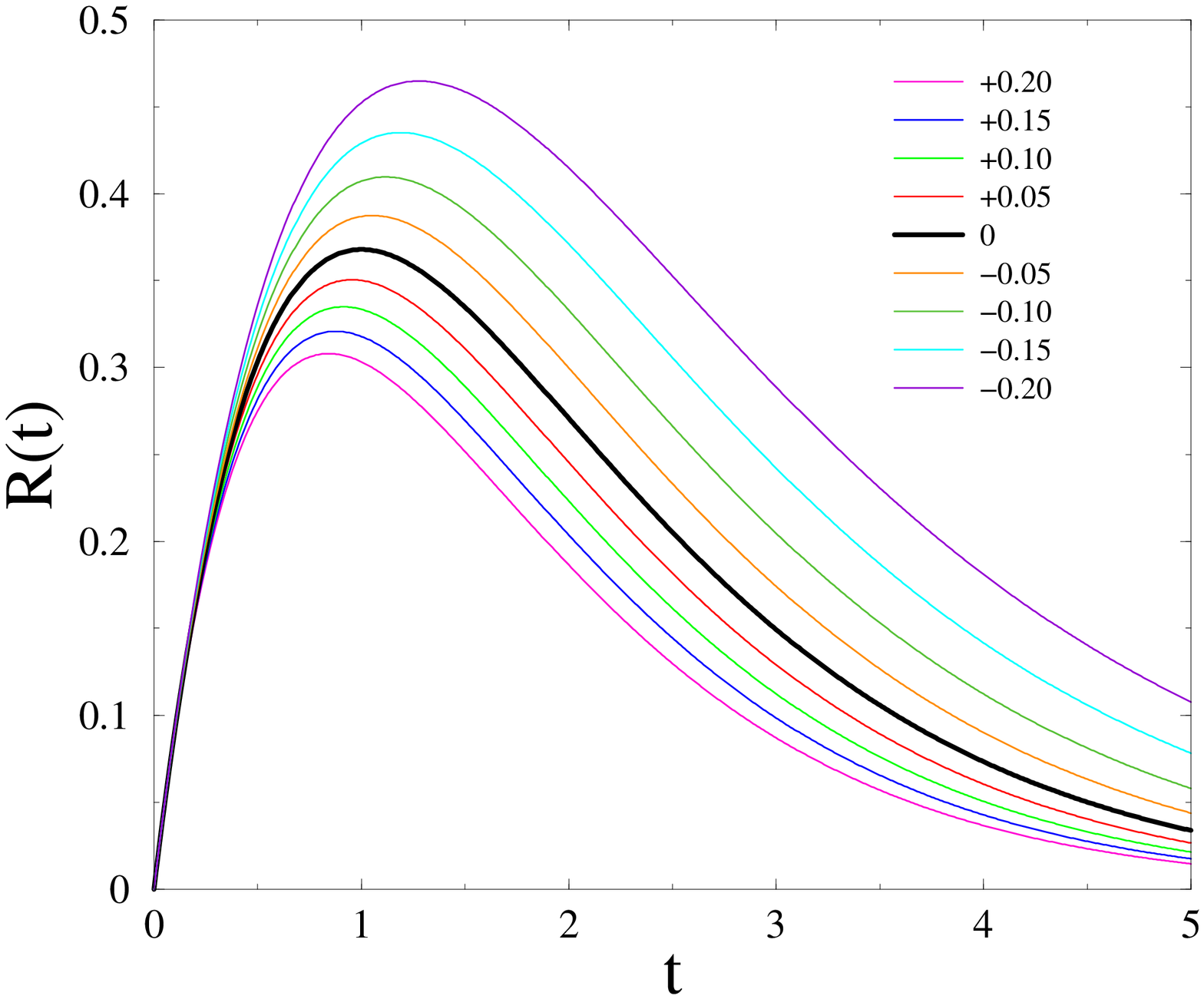}
{\hskip 10pt}
\includegraphics[angle=-90,width=.45\linewidth]{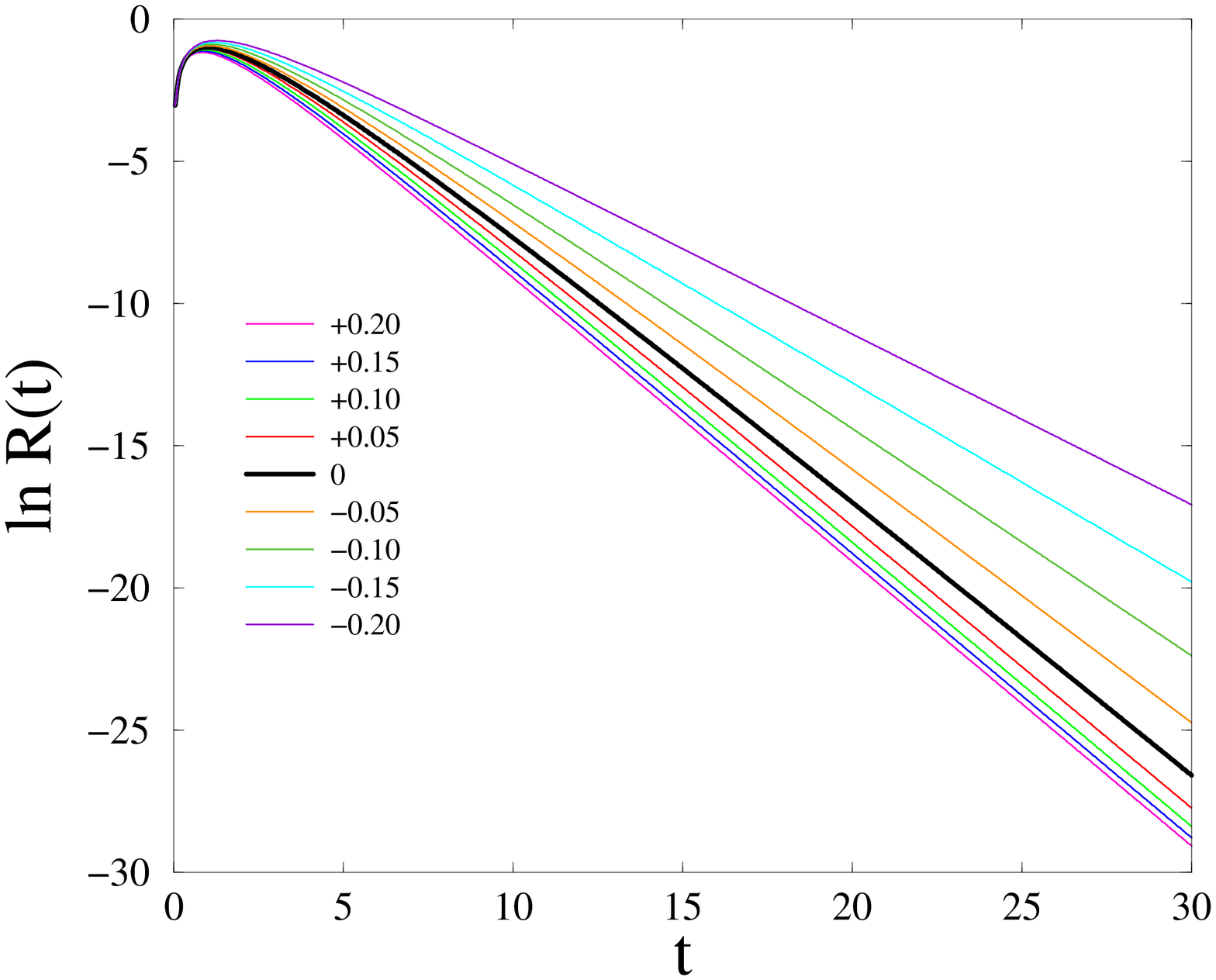}
\caption{\label{fig:resp} Plot of the response function $R(t)$ against time $t$ for various values of $\delta$\@. Left: linear scale. Right: logarithmic scale, over a larger range of times.\labeld{fig:resp}}
\end{center}
\end{figure}

The response function increases from zero, reaches a maximum, and falls off to zero, irrespective of $\delta$\@. The integrated response,
\begin{equation}
\rho=\int_0^\infty R(t)\,\d t=\frac{1}{1+2\delta}=\frac{1}{2w},
\end{equation}
decreases smoothly as a function of $\delta$, just as the response function at any fixed time $t$\@. The resonance phenomenon at $\delta=0$ does not yield any particular attribute for the response function. This point can however be shown to demarcate between two different regimes of asymptotic decay: $R(t)$ indeed falls off as $\e^{-t}$ all over the range $\delta>0$, i.e., $w>\half$, whereas its decay is slower, as $\e^{-(1+2\delta)t}=\e^{-2wt}$, for $-\half<\delta<0$, i.e., $0<w<\half$\@. These features are illustrated in
Figure~\ref{fig:resp}, showing plots of the response function $R(t)$ against time $t$, both on a linear scale (left) and on a logarithmic scale over a larger range of times (right), for the same values of $\delta$\@.

\section{Complex `fish bone' spectra in the non-integrable case \label{sec:nongeneric}\labeld{sec:nongeneric}}

Our earlier construction showing that the Markov matrix has an integrable spectrum with large degeneracies, is known to fail in the following circumstance: the parameter $\eps$ is special (i.e., $\eps=p/q$ with $p+q\le N-1$) and the zero-field rate reads $w=\half+\delta$ with $\delta\neq0$\@.

In this circumstance, the spectrum of the Markov matrix is indeed significantly different. Let us consider $\eps=1$ and free boundary conditions for definiteness.

The first occurrence of a non-integrable spectrum is at $N=3$\@. The Markov matrix has the following 8 eigenvalues:
\begin{eqnarray}
&&\lambda=0,\qquad \lambda=-1,\qquad
\lambda=-1-2\delta,\qquad \lambda=-2-2\delta,\nonumber\\
&&\lambda=-2-2\delta\pm\sqrt{1+\delta+2\delta^2},\nonumber\\
&&\lambda=-2-2\delta\pm\sqrt{\delta(2\delta-1)}.
\label{eq:fishn3}
\end{eqnarray}
\labeld{eq:fishn3}
The last 2 eigenvalues are complex for $0<\delta<\half$, i.e., $\half<w<1$\@.

Similarly, for $N=4$, the 16 eigenvalues of the Markov matrix read:
\begin{eqnarray}
&&\lambda=0,\qquad \lambda=-1\hbox{ (mult.~2)},\qquad \lambda=-2,\nonumber\\
&&\lambda=-1-2\delta,\qquad \lambda=-2-2\delta\hbox{ (mult.~2)},\qquad \lambda=-3-2\delta,\nonumber\\
&&\lambda=-2-2\delta\pm\sqrt{1+\delta+2\delta^2},\nonumber\\
&&\lambda=-3-2\delta\pm\sqrt{1+\delta+2\delta^2},\nonumber\\
&&\lambda=-2-2\delta\pm\sqrt{\delta(2\delta-1)},\nonumber\\
&&\lambda=-3-2\delta\pm\sqrt{\delta(2\delta-1)}.
\label{eq:fishn4}
\end{eqnarray}
\labeld{eq:fishn4}
The last 4 eigenvalues are complex for $0<\delta<\half$, i.e., $\half<w<1$\@. The spectra however keep the inclusion property: the 16 eigenvalues~(\ref{eq:fishn4}) include the 8 eigenvalues~(\ref{eq:fishn3}).

\begin{figure}[!ht]
\begin{center}
\includegraphics[angle=-90,width=.45\linewidth]{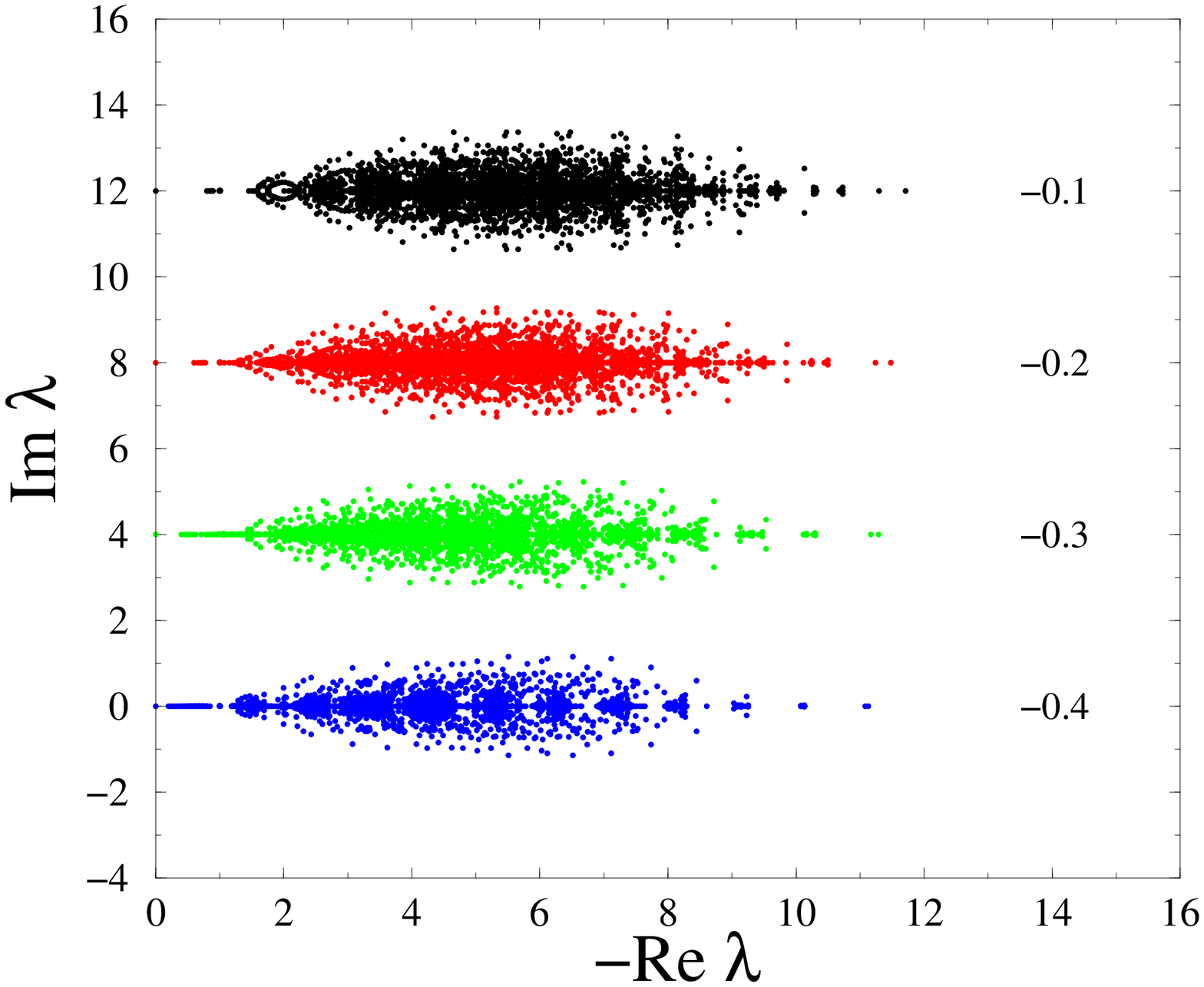}
{\hskip 10pt}
\includegraphics[angle=-90,width=.45\linewidth]{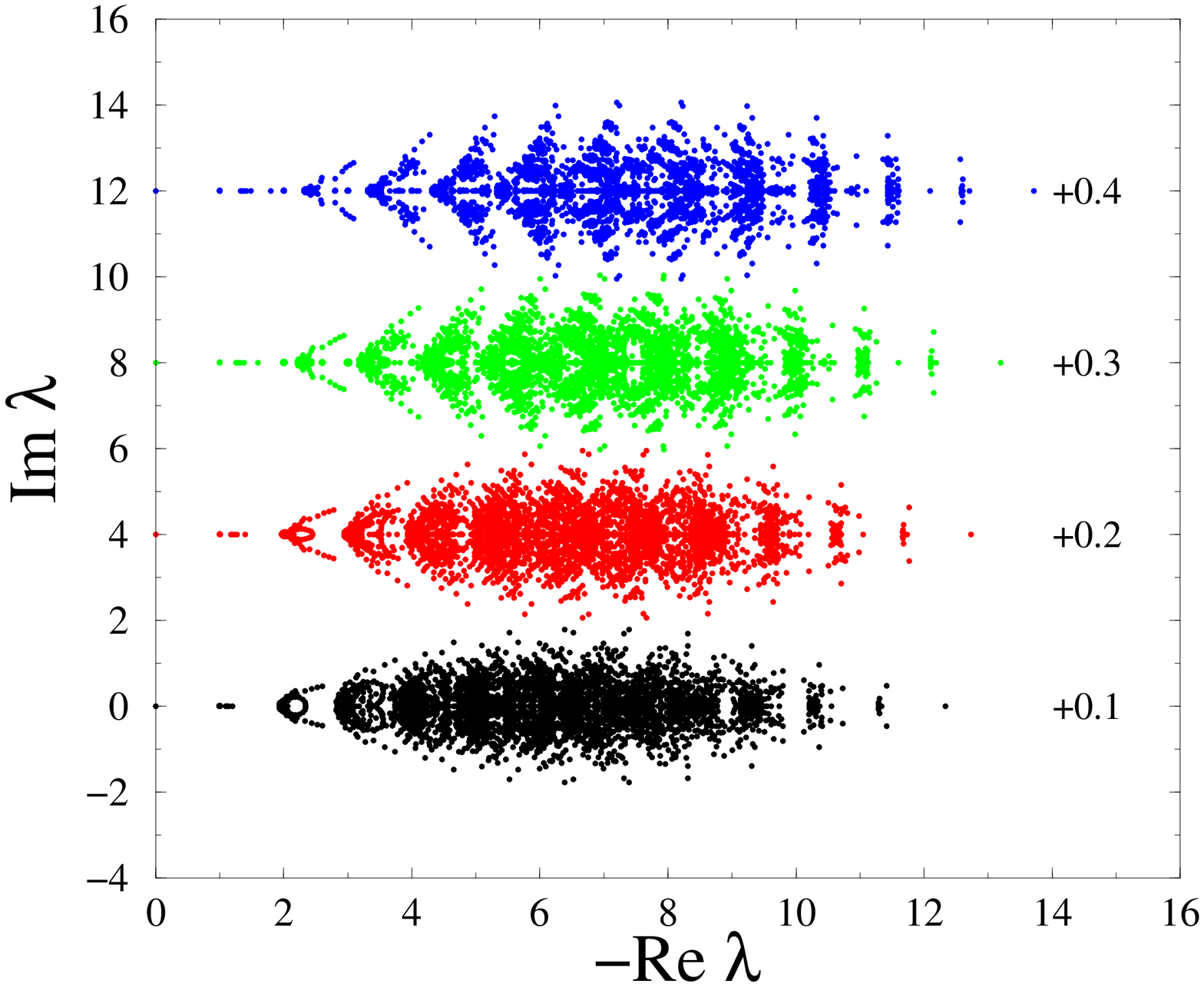}
\caption{\label{fig:fish} Plot of the spectra of the Markov matrices in the complex $\lambda$ plane, for $N=12$ and various values of $\delta$ (which are indicated to the right of the corresponding plot). Each spectrum is symmetric with respect to the real axis and consists of $2^{12}=4096$ points. Spectra are translated vertically and shown in various colors for clarity. Left: $\delta<0$\@. Right: $\delta>0$\@.\labeld{fig:fish}}
\end{center}
\end{figure}

As the system size $N$ grows, more of the combinatorial degeneracies of the integrable case are lifted, and more eigenvalues become complex. These features are illustrated in Figure~\ref{fig:fish}, showing the spectrum of the Markov matrix in the complex $\lambda$ plane, obtained by means of a numerical diagonalization, for $\eps=1$ and $N=12$, with free boundary conditions and for various values of $\delta$\@. The spectra progressively become structured as `fish bones' as $|\delta|$ increases.

\begin{figure}[!ht]
\begin{center}
\includegraphics[angle=270,width=.45\linewidth]{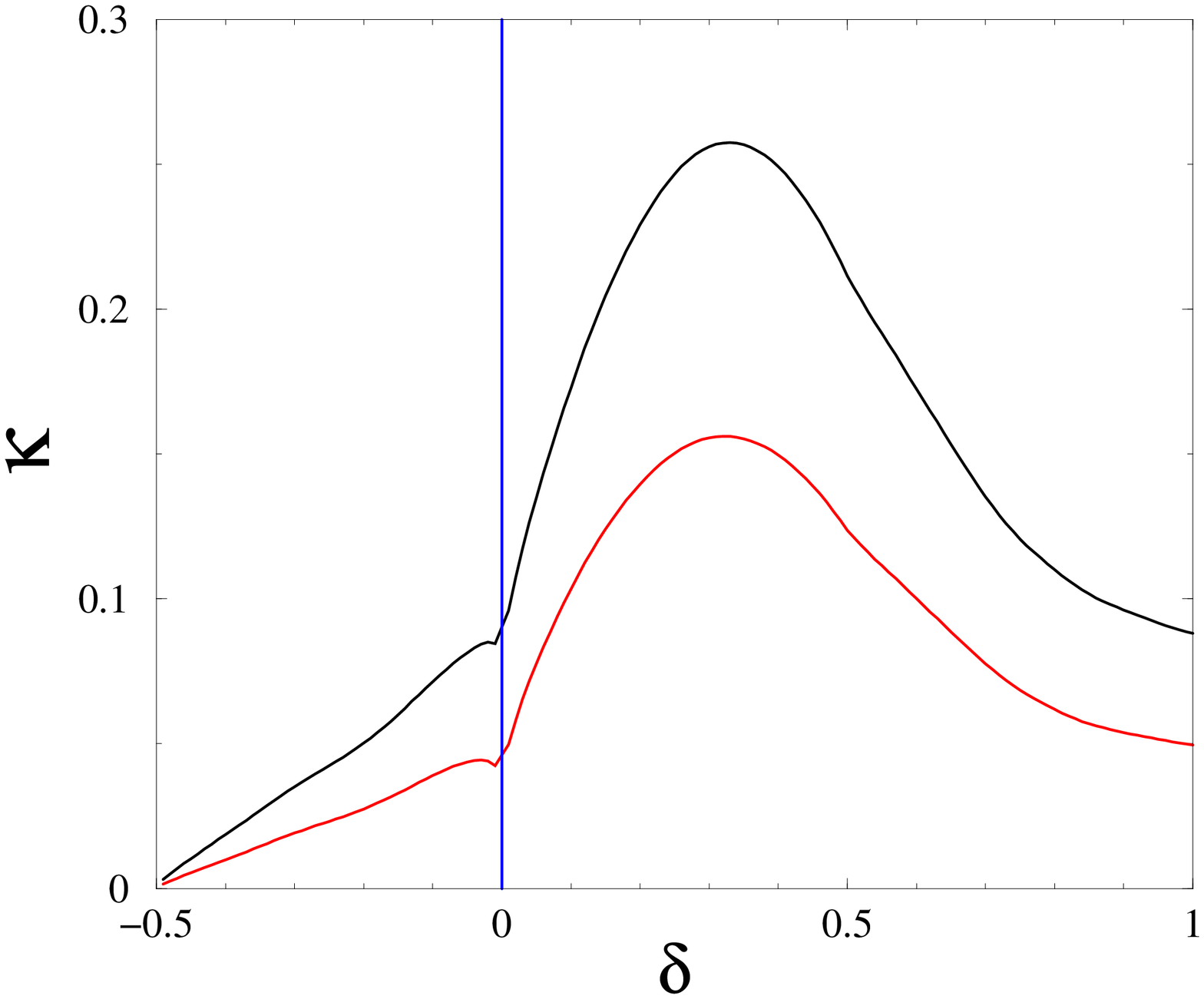}
\caption{\label{fig:im2} Plot of the mean squared imaginary part, $\kappa$, of the spectrum of the Markov matrix against $\delta=w-\half$ for $\eps=1$ and $N=10$\@. Upper black curve: free boundary conditions. Lower red curve: fixed boundary conditions. Vertical blue line: integrable case ($\delta=0$).\labeld{fig:im2}}
\end{center}
\end{figure}

A quantitative measure of the extension of the fish bone spectra in the complex plane is provided by the mean squared imaginary part
of the spectrum,
\begin{equation}
\kappa=\mean{(\Im\lambda)^2}=\frac{1}{2^N}\sum_{a=1}^{2^N}(\Im\lambda_a)^2.
\label{eq:im2def}
\end{equation}
\labeld{eq:im2def}
This quantity is plotted in Fig.\ \ref{fig:im2} against $\delta=w-\half$ for $\eps=1$, $N=10$, and free and fixed boundary conditions. Data have been obtained by means of a numerical diagonalization of the Markov matrix for values of $\delta$ on a grid with mesh 0.01\@. The integrable case ($\delta=0$) is shown as a vertical blue line. The data suggest that $\kappa$ does not vanish in the vicinity of that point. In other words, the limits $N\to\infty$ and $\delta\to0$ do not commute. The quantity $\kappa$ rather exhibits a cusp (change of slope) at $\delta=0$, whereas the observed dip is most certainly a finite-size effect. Finally, the mean squared imaginary part of the spectrum has a smooth maximum near $\delta\approx0.3$, i.e., $w\approx0.8$, irrespective of boundary conditions.

\subsection{Sensitivity: fish bones and false diagonalization\label{sec:falsediag}\labeld{sec:falsediag}}

In this section we study two related phenomena: for non-generic $\eps$, the rapidity with which the imaginary parts of the eigenvalues grow (as a function of $\delta$), as exemplified in the ``fish bone'' graphs of Fig.\ \ref{fig:fish}, and another anomaly, mentioned in Remark \ref{rem:isospectraltwo}, connected to numerical diagonalization in the case that a Jordan form \textit{is} needed. What happens is that the computer spits out bona fide eigenvectors and non-degenerate eigenvalues, usually with small complex parts. The complex parts typically are much larger than the rounding error of the software. (This can also fool the investigator: if you get an imaginary part of order $10^{-3}$ and if your accuracy is $10^{-16}$, then you tend to believe the numerics. In our case we had analytic proof that the spectrum (with zero $\delta$) consisted of real non-positive integers.)

What we show in this section is how, for Jordan forms, their non-generic nature expresses itself in vastly exaggerated effects of rounding errors, including imaginary parts for real eigenvalues and rapid growth of the imaginary part when true perturbations move the spectrum off the real line.

Consider the archetypal matrix requiring the Jordan form
\be
R\equiv\lsmatrix{cccccccc}
{   0  & 1 & 0 & 0 & 0      & 0 & 0  & \dots \\
    0  & 0 & 1 & 0 & 0      & 0 & 0  & \dots \\
       &   &   &   & \ddots &   &    &  \\
    0  & 0 & 0 & 0 & 0      & 0 & 0  & 1 \\
    0  & 0 & 0 & 0 & 0      & 0 & 0  & 0
}
\,,
\label{eq:R}
\ee
that is, the $n$-by-$n$ matrix, $R$, has non-zero entries only for $R_{k\ell}=\delta_{k,\ell+1}$\@. Of course this is already in Jordan form. Consider the effect of a single non-zero entry added somewhere to $R$\@. This could correspond to a rounding error (say of order $10^{-16}$) or for the fish bone calculation this would be $\delta$\@. The additional term is $xM$, where $x$ is a number and $M$ is a matrix, all zeros, except for a single 1, whose location is described below\@. There are several cases.
\begin{enumerate}
\item $M$'s non-zero entry is on the diagonal. Then  $\det\left(\lambda\one-R-xM\right)=\lambda^{n-1}(\lambda-x)$\@. This implies that the shift in eigenvalue is just $x$, linear in the change.
\item $M$'s non-zero entry is anywhere above the diagonal. Then $\det\left(\lambda\one-R-xM\right)=\lambda^{n}$, and there is \textit{no} change in the eigenvalue.
\item $M$'s non-zero entry is on the $j^\mathrm{th}$ sub-diagonal, that is $M_{k\ell}=1$ for $k-\ell=j$, $n-1\ge j>0$\@. Then $\det\left(\lambda\one-R-xM\right)=\lambda^{n-j-1}\left(\lambda^{j+1} -x\right)$\@. The non-zero eigenvalues are then $\{|x|^{1/(j+1)}\omega\}$ where $\omega$ is one of the $(j+1)^\mathrm{th}$ roots of unity.\label{item:belowdiag}
\end{enumerate}
The extreme situation occurs in Item \ref{item:belowdiag}, when $x$ sits in the $(n,1)$ position, leading to $n^\mathrm{th}$ roots of unity. For $n>2$ these necessarily include complex values. For $n=2$ the deviation can be real only. In any case it's clear how errors enormously larger than $10^{-16}$ can occur. If there is a numerical error of the sort involving a Jordan block arising from degeneracy 6 (which happens for $\ge4$ spins) there can easily be complex roots of order $10^{-3}$\@. In an ordinary numerical calculation one would be hard put to call this a rounding error.

In practice one does not start with a matrix of the form \Eqref{eq:R}, but in the course of numerical diagonalization the many similarity transformations can move deviations of the pure Jordan form (whether intentional ($\delta$) or not (rounding error)), anywhere, including dangerous spots, like the $(n,1)$ position. Thus suppose the true Jordan form is achieved by the transformation $w=VJV^{-1}$ with $J$ the Jordan form and $V$ the similarity transformation. Then a slightly modified $w$ subjected to this same transformation will have small non-zero terms in ``dangerous'' locations, as described above.

\section{Sum rule for the Markov matrix\label{sec:sumrule}\labeld{sec:sumrule}}

In this section we provide an alternative investigation of the spectrum of the Markov matrix. We employ an explicit calculation of the average absolute eigenvalue
\begin{equation}
\Lambda=-\mean{\lambda}=-\frac{1}{2^N}\sum_{a=1}^{2^N}\lambda_a=-\frac{\tr w_N}{2^N}.
\label{eq:sumdef}
\end{equation}
\labeld{eq:sumdef}
This quantity can be evaluated by elementary means, thereby providing a sum rule for the spectrum. We have
\begin{equation}
\Lambda=\frac{1}{2^N}\sum_{\C=1}^{2^N}\omega(\C),
\end{equation}
where $\C=\{\s_1,\dots,\s_N\}$ is an arbitrary spin configuration, and $\omega(\C)$ denotes the total exit rate from that configuration.

Let us consider first free boundary conditions ($h_1=0$). For $\eps$ generic, and for any configuration $\C=\{\s_1,\dots,\s_N\}$,
flipping the uppermost spin $\s_1$ always brings a contribution $w=\half+\delta$ to $\omega(\C)$, whereas the contribution coming from
flipping any other spin ($\s_n$ for $n=2,\dots,N$) is 1 if $\s_n\ne\sign h_n$ and 0 if $\s_n=\sign h_n$\@. We thus obtain
\begin{equation}
\Lambda=\frac{N}{2}+\delta.
\label{eq:sumgen1}
\end{equation}
\labeld{eq:sumgen1}
For $\eps=p/q$ special, the contribution to $\omega(\C)$ coming from flipping $\s_n$ is also $w=\half+\delta$ whenever $h_n=0$\@. The
number of configurations $\C$ such that $h_n=0$ at depth $n=k(p+q)+1$ reads $2^NP_k$, where
\begin{equation}
P_k=\frac{\cc{k(p+q)}{kq}}{2^{k(p+q)}}\qquad(k\ge0)
\end{equation}
is the probability for a random walker on the line, making integer steps $+p$ and $-q$ with equal probabilities, to be back to its
starting point after $k(p+q)$ steps. The result~(\ref{eq:sumgen1}) is thus changed to
\begin{equation}
\Lambda=\frac{N}{2}+A_N\delta,
\label{eq:sumspec1}
\end{equation}
\labeld{eq:sumspec1}
where for all $N$ in the range $k(p+q)+1\le N\le(k+1)(p+q)$,
$A_N$ is constant and equal to
\begin{equation}
\A_k=\sum_{l=0}^kP_l.
\end{equation}

Let us now consider fixed boundary conditions ($\s_0=+1$, $h_1=-1$). For $\eps$ generic, the contribution to $\omega(\C)$ coming from
flipping any spin $\s_n$ is 1 if $\s_n\ne\sign h_n$ and 0 if $\s_n=\sign h_n$\@. We thus obtain
\begin{equation}
\Lambda=\frac{N}{2}.
\label{eq:sumgen2}
\end{equation}
\labeld{eq:sumgen2}
For $\eps=p/q$ special, the contribution to $\omega(\C)$ coming from flipping $\s_n$ is $w=\half+\delta$ whenever $h_n=0$\@. The number
of configurations $\C$ such that $h_n=0$ at depth $n=k(p+q)$ reads $2^NQ_k$, where
\begin{equation}
Q_k=\frac{\cc{k(p+q)-1}{kq}}{2^{k(p+q)-1}}=\frac{2p}{p+q}P_k\qquad(k\ge1)
\end{equation}
is the probability for a random walker on the line, making integer steps $+p$ and $-q$ with equal probabilities, and starting from
$-q$, to be at the origin after $k(p+q)-1$ steps. The result~(\ref{eq:sumgen2}) is thus changed to
\begin{equation}
\Lambda=\frac{N}{2}+B_N\delta,
\label{eq:sumspec2}
\end{equation}
\labeld{eq:sumspec2}
where for all $N$ in the range $k(p+q)\le N\le(k+1)(p+q)-1$,
$B_N$ is constant and equal to
\begin{equation}
\B_k=\sum_{l=1}^kQ_l=\frac{2p}{p+q}(\A_k-1).
\end{equation}

The above results are summarized in the last column of Table~\ref{table:spectra}. In all the integrable situations where the spectrum is known explicitly, the corresponding simple values of $\Lambda$ can be checked directly. The special results~(\ref{eq:sumspec1}) and~(\ref{eq:sumspec2}), with their non-trivial amplitudes $A_N$ and $B_N$, give some information on the non-integrable situations, namely the position of the center of mass of the corresponding fish bone spectra.

The generic results~(\ref{eq:sumgen1}) and~(\ref{eq:sumgen2}) can be recovered from the special ones~(\ref{eq:sumspec1}) and~(\ref{eq:sumspec2}) by taking the limit of an infinite period ($p+q\to\infty$), so that $A_N=\A_0=1$ and $B_N=\B_0=0$ for all system sizes $N$\@. Furthermore, if the integers $p$ and $q$ are interchanged, so that $\eps$ is changed to its inverse, the $P_k$ and the $A_N$ are left
invariant, whereas the $Q_k$ and the $B_N$ are multiplied by $q/p$\@.

The case where $\eps=1$, i.e., $p=q=1$, is that of the usual symmetric walker. The probabilities
\begin{equation}
P_k=Q_k=\frac{\cc{2k}{k}}{2^{2k}}
\end{equation}
have a slow power-law decay, as $P_k\approx(\pi k)^{-1/2}$\@. As a consequence, the amplitudes $A_N$ and $B_N$ grow as
\begin{equation}
A_N\approx B_N\approx\left(\frac{2N}{\pi}\right)^{1/2}.
\end{equation}
The contributions proportional to $\delta$ in~(\ref{eq:sumspec1}) and~(\ref{eq:sumspec2}) grow indefinitely, but sub-extensively, and therefore remain negligible with respect to the leading terms $N/2$\@.

In all the other rational cases, i.e., $\eps=p+q$ (irreducible) with $p+q\ge3$, the walker is biased, and so the return probabilities decay exponentially, as $P_k\sim Q_k\sim a^k$, with
\begin{equation}
a=\frac{(p+q)^{p+q}}{2^{p+q}p^pq^q}.
\end{equation}
As a consequence, the amplitudes $A_N$ and $B_N$ saturate to finite limits:
\begin{equation}
A_\infty=\sum_{k\ge0}P_k,\qquad
B_\infty=\frac{2p}{p+q}(A_\infty-1).
\end{equation}
These limits turn out to be algebraic numbers. This property stems from the fact that the return probabilities $P_k$ and $Q_k$ are related to the Fuss-Catalan numbers \cite{graham}. Some variants of the above series have been met in several works in the physics literature \cite{derridaPRL1998, derridaJSP1999, bauerJSP1999}.

To close with an example, for $\eps=2$ we obtain $A_\infty=1+3/\sqrt{5}$ and $B_\infty=4/\sqrt{5}$, whereas for $\eps=1/2$ we obtain
the same $A_\infty$, as expected, but $B_\infty=2/\sqrt{5}$\@.

\section{Bringing the stochastic generator to triangular form \label{sec:triangular} \labeld{sec:triangular}}

\font\capsten=cmcsc10
\font\capseight=cmcsc8
\def\matlab{{\capseight MATLAB}}

In Remark \ref{rem:y} (in Sec.\ \ref{sec:recursion}) we defined the matrix $y$ (or $y_N$) associated with a random walk on the edges of an $N$-cube and presented its full spectral analysis. Its basis vectors, in Sec.\ \ref{sec:spincorrelations}, acquired the interpretation of correlation functions, and we saw that in that basis the master equation has a triangular form. In this section we briefly derive that triangular form using the algebraic-recursion relation approach, noting as a consequence that its validity extends beyond the zero-temperature granular dynamics model.

Let $\{V_\alpha\}$ be the set of eigenvectors of $y$\@. Writing these vectors as the columns of a matrix $V_{\cdot\alpha}$, we have $\sum_\ell y_{k\ell}V_{\ell\alpha}=\lambda_\alpha V_{k\alpha}$\@. The state numbering, and resultant labeling of $V$, follows the conventions for the matrix $w$\@. Then the matrix $\widetilde w\equiv V^{-1}wV$ is lower triangular. With this numbering, neither the eigenvalues of $y$ nor those of $w$ come in descending order.

\noindent\textsf{Proof}:~
We first note the following recursion for the diagonalizing matrix for $y$\@. N.B. this uses the state-ordering scheme outlined in Sec.~\ref{sec:w} and is a direct consequence of the eigenvector construction.
\be
V_{N+1}=\left(
         \begin{array}{rrr}
         V_N&~&V_N\\
         V_N&&-V_N
         \end{array}
         \right)
\label{eq:Vrecursion}
\ee
\labeld{eq:Vrecursion}~\\
Moreover, it has the following properties:
\bea
V_N^\top&=&V_N   \label{eq:Vtranspose} \\
V_N^{-1}&=&V_N/2^N  \label{eq:Vinverse}
\eea
Recall too
\be
w_{N+1}=
\lsmatrix{ccc}
{w_N-\tilde\Delta &~& \Delta \\
 \tilde \Delta && w_N-\Delta}
\,.
\label{eq:wnplusone}
\ee
We suppose that for $N$, $\widetilde w_N\equiv V_N^{-1}wV_N$ is lower triangular. This is the inductive hypothesis. It is true for $N=1$ since in this case $w$ coincides with $y$ and is diagonalized by the transformation.

Consider
\be
\VNplusinv w_{N+1} V_N = \frac1{2^{N+1}}
                          \left(
                           \begin{array}{rrr}
                           V_N &~& V_N \\
                           V_N && -V_N
                           \end{array}
                           \right)
                           \lsmatrix{ccc}
                           {w_N-\tilde\Delta &~& \Delta \\
                           \tilde \Delta && w_N-\Delta}
                           \left(
                           \begin{array}{rrr}
                           V_N &~& V_N \\
                           V_N && -V_N
                           \end{array}
                           \right)
\ee
Performing the multiplications and making use of \Eqref{eq:Vinverse} yields
\bea
\widetilde w_{N+1}&=&\VNplusinv w_{N+1} V_N = \frac1{2^N}
                          \left(
                           \begin{array}{rrr}
                           V_N w_N V_N &~& 0 \\
                           -\one + 2 V\Delta V && V_Nw_NV_N-\one
                           \end{array}
                           \right)     \nonumber \\
                           &=&
                           \left(
                           \begin{array}{rrr}
                           \widetilde w_{N} &~& 0 \\
                           -\one+ 2^{-(N-1)}V\Delta V && \widetilde w_{N}-\one
                           \end{array}
                           \right)
\eea
This expression is lower triangular. The exact form of $\Delta$ only enters in the lower left. This shows incidentally that all that is necessary for triangularity is the indicated recursion, recalling that there is also the condition $\Delta+\tilde\Delta=\one$\@.

\refstepcounter{remark} \label{rem:twotriangles}
\smallskip\noindent\textsf{Remark \arabic{remark}}:~
To obtain full agreement between the triangular matrix derived here and that obtained from the correlations, one must multiply each $V_\alpha$ (as defined in Remark \ref{rem:y}) by $(-1)^{|A|}$, with $A$ the subset of $\{1,\dots,N\}$ associated with the index~$\alpha$\@. This is a unitary transformation of~$V$\@.

\section{Discussion\label{sec:discussion}\labeld{sec:discussion}}

Even with random shaking (positive ``temperature''), the movement of grains has structures and constraints not present in ordinary liquids. Without shaking, the zero-temperature dynamics has yet more restrictions and we have found the mathematical expression of this feature to be reflected in the matrix generator of the stochastic dynamics in the following way: for generic values of the shape parameter, $\eps$, the matrix has degenerate eigenvalues for which the number of eigenvectors is fewer than the degeneracy: it is \textit{not} diagonalizable and the nearest one can get to a spectral expansion is the Jordan canonical form.

In this article we have presented extensive information on the properties of the generator of the stochastic dynamics (which we call $w$). The mere fact that it obeys a particular recursion relation (our \Eqref{eq:recursion}) already fixes its eigenvalues and the multiplicity of the associated invariant spaces (\textit{invariant space} is what replaces the notion of eigenvector span when they do \textit{not} span). Moreover, the recursion is more general than the granular model in the sense that other matrices also satisfy the recursion (they need not even be stochastic) and have the same spectrum. Nevertheless, this same spectrum was derived in a way that made direct use of the model's features by studying correlation functions among the spins. This approach made use of a fundamental asymmetry in the model: a given spin is affected only by those above it. The triangular structure of the differential equations obtained in this way is identical that obtained from a particular similarity transform applied to the stochastic generator. We also show that the recursion property alone is enough to guarantee that this transform puts $w$ into (lower) triangular form.

The particular similarity transform casts light on one of the lovely features of the model, namely the appearance of combinatorial coefficients for the multiplicity of the invariant spaces. In particular, the eigenvalues of $w$ are $\{0,-1,-2,\dots,-N\}$ with $N$ the number of spins in the vertical column that models the granular material. The multiplicity of eigenvalue $-k$ is $C^N_k$ (we take $\delta=0$ in this discussion)\@. Now the invariant spaces are \textit{not} characterized by the number of spins in one direction or another, which would have been a natural way to obtain combinatorial coefficients of the sort indicated. However, the triangulating transformation does count spins in the following sense. It is the diagonalizing matrix of ``$y$,'' described in Remark \ref{rem:y}. That matrix also describes $N$-spin stochastic dynamics, but for it, \textit{all} single spin flips are allowed, in contrast to the granular model with its highly structured rules for transitions. The eigenvectors of $y$  of eigenvalue $-k$ (and it \textit{has} a full complement, unlike $w$) are obtained by looking at the subsets of size $k$, so indeed there will be $C^N_k$ of them for each $k$\@. These subset-based eigenvectors are what triangularize~$w$\@.

The relation of the eigenvalues and the combinatorial coefficients is also clear in the development of the differential equations for the correlation functions (Sec.\ \ref{sec:spincorrelations}). When each spin can flip independently, the combinatorial coefficients, as well as the full spectrum, can be obtained by looking at subsets of the spins of particular cardinality---equivalent to the diagonalization of the generator of the random walk on the hypercube (what we called $y$). For the model discussed here we do have dependence on the motion of other spins, but it is asymmetric, with a given spin's dynamics depending only on those above it. This is sufficient to imply that the differential equations for the correlations have a triangular form, and that both the eigenvalues and the cardinality of the invariant spaces is unchanged from the all-independent model. Finally, in a satisfying confluence of the two approaches, all coefficients in the triangular form obtained in this way coincide with those of the pure matrix approach.

Our mathematical developments have not completely answered, ``why Jordan?'' nor the yet more compelling question, ``why maximally Jordan?'' Certainly this property is a reflection of the many constraints in the system. These constraints are in their own way ``maximal,'' in the sense that for every potential spin flip at level $n$, \textit{all spins above} $n$ are queried. It is thus plausible that this maximal constraint on configurations induces the cascading structure in the ``maximally Jordan'' invariant spaces. In the time domain, a consequence of the Jordan form is that the relaxation process is not direct, and we have explored the way in which the Jordan form induces non-monotonic behavior, despite the presence of entirely real spectrum (for non-generic $\eps$ and nonzero $\delta$ complex eigenvalues obviously allow non-monotonicity). In Sec.\ \ref{sec:maxjordan} we gave a mathematical criterion for the maximal Jordan property and from the correlation function approach the interdependence and triangular form of the matrix suggests this property. But it must be cautioned that the recursion alone, or triangularity alone does not compel the maximal Jordan property and we have produced examples (not presented in this article) in which various patterns of the matrix $\Delta_N$ (diagonal and with 0's 1's and $1/2$'s on its diagonal), when used in the recursion, do \textit{not} lead to the maximal Jordan property. Nevertheless, at the heuristic level our picture is that a cascade of processes takes place, with earlier transitions needed before later ones can happen. This is different from what happens in a (say) spin glass where individual configurations must do unlikely things in order to move to lower energy states. In the grain model it is the probability distribution that behaves non-monotonically, which is not the case for the spin glass. However, the ``cascading'' is in a way less dramatic than what happens in the spin glass. There, arriving at a lower energy state can take exponentially long; for us, the delay only slows things down in a polynomial way, with the dominant (and not particularly slow) exponential ultimately manifesting itself. Taking heed of this example, we note that ``hindered,'' ``complex dynamics'' and ``Jordan forms'' are independent concepts. In particular, most models exhibiting glassy dynamics and aging have explicit reference to a static Hamiltonian. The resulting stochastic dynamics therefore obeys detailed balance. The associated Markov matrix can be brought to a symmetric form. It is therefore diagonalizable, and its spectrum is real. These features are best exemplified by the so-called ``kinetically constrained'' models \cite{garrahan}, such as the Frederickson-Andersen spin model \cite{ritort}, where entropic barriers in the configuration space place restrictions on allowed transitions, but, as indicated, do not require the Jordan form. As a corollary, in these models the time dependence of the \textit{probability distribution} is always monotonic in its approach to the stationary (equilibrium) state, a property we have explicitly shown not to hold for our model.

The principal results of the present work are concerned with the short-time relaxation of the model, dealing as they do with eigenvalues and with the huge---for large magnitude eigenvalues---invariant spaces of the Markov matrix. This does not deal with many other physical issues, many of which are concerned with the structure of the stationary state. Thus many of the short time properties we have found hold irrespective of boundary conditions (see Table~\ref{table:spectra}), whereas the essential physical attributes of the model \textit{are} sensitive to this feature \cite{luckmehta2, mehtaluck4, luckmehta1, luckmehta3}. For free boundary conditions (the situation implicitly under consideration until Sec.\ \ref{sec:general}) the stationary state is always a non-trivial fluctuating non-equilibrium steady state. For fixed boundary conditions (e.g., $\s_0=+1$), the nature of the stationary state depends on parameters. For $\eps$ generic (i.e., essentially irrational), the system is driven to a unique ground state. The ground-state configuration is absorbing for the zero-temperature dynamics. For an infinite system, it is a quasiperiodic sequence, which admits a geometrical construction. For $\eps$ special (i.e., essentially rational), so that there are internal points of zero field, the system reaches a fluctuating stationary state characterized by an anomalous scaling of fluctuations.

\begin{acknowledgments}
We thank Deepak Dhar, Bernard Gaveau, and Leonard J. Schulman for helpful discussions. AM and LSS are grateful for the hospitality of the Institut de Physique Th\'eorique where much of this work was done.
\end{acknowledgments}

\appendix

\section{Conjugation transformation \label{sec:conjugation}\labeld{sec:conjugation}}

\def\nbyn#1{#1 \mskip-4mu \times\mskip-4mu #1}

In this Appendix we perform---at the matrix level---the transformation carried out on the determinant in Eqs.~(\ref{eq:det-step0})--(\ref{eq:detmanipulations}). It is a systematic procedure, different from that appearing in Sec.\ \ref{sec:triangular}, for obtaining a triangular form of the matrix.

As above, $\one_k$ and $\zero_k$ are the $2^k\!\!\times\!2^k$ unit and zero matrices, respectively. Let $a_k\equiv\lsmatrix{rr}{\one_k&\one_k\\ \zero_k&\one_k}$, with inverse $a_k^{-1}=\lsmatrix{rr}{\one_k&-\one_k\\ \zero_k&\one_k}$\@. Now fix $N$ (the number of spins). We next define $A_k^{(N)}$ (or $A_k$, with the $N$ suppressed). For this $N$, concatenate $2^{N-k}$ $a_k$'s along diagonal blocks so as to make a $\nbyn{2^N}$ matrix. This concatenated object is $A_k^{(N)}$\@. Thus for $k=1$, $a_1$ is a $\nbyn2$ matrix, so that $A_1$ has $2^{N-1}$ such diagonal blocks:
\be
A_1=\lsmatrix{ccccc}
{a_1   & \zero_1  & \dots  &\dots & \zero_1\\
 \zero_1   & a_1  & \dots  \\
       &      & \ddots \\
 \zero_1   & \zero_1  & \dots  &      &a_1}
\,.
\ee
For given $N$, the largest ``$A_k$'' is $A_{N-1}$ (which is composed of two copies of $a_k$). The inverse of $A_k$ is constructed in the same way from the inverses of $a_k$\@. If $Q=\lsmatrix{rr}{\alpha&\beta\\ \gamma&\delta}$, then $P=a\,Q\,a^{-1}=\lsmatrix{rrr}{\alpha+\gamma&~&-\alpha-\gamma+\beta+\delta\\\gamma&&-\gamma+\delta}$\@. For our matrices, either $w_N$ or $w_N-\lambda\one_N$, the upper right block of $P$ is zero, so that a succession of $A$ transformations can bring $w$ to lower triangular form. The procedure for doing this is to work your way down: First transform $w_N$ by $A_{N-1}$ then by $A_{N-2}$, and so forth, with the final transformation using $A_1$\@. What then sits on the diagonal are the eigenvalues of $w$\@.

This technique works whether or not a Jordan form is required. Thus for the fully connected graph the recursion that yields the continuous time generator uses $\Delta=(1/2)\bm 1$ at each stage of the construction. As for $w$, the associated matrix is disassembled by the $A$'s, but since there are $2^N$ eigenvectors, no Jordan form is needed.

\section{Quasisymmetries\label{sec:quasisymmetry}\labeld{sec:quasisymmetry}}

\subsection{The operator $F$}

Recall that the (diagonal) matrix in the \ll\ block of $w$ is $\tilde \Delta$ (see \Eqref{eq:recursion}). Although it might seem that the 0 and 1 count should be the same for both it and $\Delta$, there is nevertheless an important difference. The role of source and target is reversed and therefore the sequence is reversed. This can be formally stated by defining a family of matrices $F_N$ to be the left-right reversal of the $2^N\!\times\!2^N$ identity ($\one_N$). Thus for example
\be
F_2=\lsmatrix{cccc}{0&0&0&1\\0&0&1&0\\0&1&0&0\\1&0&0&0}
\,.
\ee
Note that $F^2=\one$\@. With this notation
\be
\tilde\Delta_N=F_N\,\Delta_N\,F_N\,, \quad \forall N
\,.
\ee

\subsection{Intertwining operators}
Let
\be
P_N=[1,1]\otimes \one_N\,,
\,
\ee
which is a mapping from $\realnumbers^{2^{N+1}}\to\realnumbers^{2^{N}}$ and (e.g.) for $N=2$ takes the form
\be
P_2=\left( \begin {array}{cccccccc}
1&0&0&0&1&0&0&0\\
0&1&0&0&0&1&0&0\\
0&0&1&0&0&0&1&0\\
0&0&0&1&0&0&0&1
\end {array} \right)
\,.\ee
Alternatively,
$
P_N^\top=\lsmatrix{c}{\one_N\\ \one_N}
$\@.
Interest in this operator arises from the identity
\be
2A_N=P_N\, A_\Np \,P_N^\top
\,
\ee
where $A_N$ can be either $w_N$ or its transpose $g_N$\@. This operator has an interpretation as a projection, mapping pairs of spin states onto a single spin state. It satisfies the identities
\be
P_N P_N^\top=2\,\one_N  \,, \qquad P_N^\top P_N =\lsmatrix{cc}{1&1\\ 1&1}\otimes \one_N
\,.
\ee
The ``doubling'' of Sec.\ \ref{sec:eigenvectors} is multiplication by~$P_N^\top$\@. Also note (which is what lies behind the demonstrations of that Section) that
\be
w_{N+1}^\top P_N^\top=\lsmatrix{c}{w_N^\top\\ w_N^\top}
\ee
since the $\Delta$'s are eliminated, as in \Eqref{eq:doubling}\@.

The operators $P$ and $F$ account for the symmetries observed in the form of the eigenvectors.

\bigskip

\noindent\textbf{References}
\vskip -.5in


\begin{thebibliography}{34}
\bibitem[{not({\natexlab{a}})}]{note:donotneedproof}
\bibinfo{note}{We do not supply a proof of irreducibility since we later derive
  the full spectrum and find the eigenvalue 0 to be nondegenerate. Since $A_0$
  vanishes nowhere this implies irreducibility.}
\bibitem[{not({\natexlab{b}})}]{note:rational}
\bibinfo{note}{For any fixed $N$, rational numbers with sufficiently large
  denominators will behave ``generically,'' in the sense that the field, $h$,
  cannot vanish.}
\bibitem[{not({\natexlab{c}})}]{note:OR}
\bibinfo{note}{The observable representation \cite{multiplephases,
  meanfieldobsrep, imaging, visualizingOR}, mentioned in the Introduction, is
  an embedding of a discrete space in a continuous one, based on spectral
  properties of matrices relating the points of the discrete space. Distances
  in the continuous space often reflect significant properties. For random
  walks the observable representation can give a clear visualization of the
  space on which the walk takes place.}
\end{thebibliography}

\end{document}